\documentclass[sigconf, screen]{acmart}
\usepackage{amsmath,amsfonts}
\usepackage{algorithm}
\usepackage{algorithmicx}
\usepackage{algpseudocode}
\usepackage{graphicx}
\usepackage{textcomp}
\usepackage[svgnames]{xcolor}
\usepackage{bm}
\usepackage{subcaption}
\usepackage{svg}
\usepackage{hyperref}
\usepackage{fancyhdr}
\usepackage{soul}
\usepackage{gensymb}
\usepackage{cleveref}
\usepackage{nicematrix}
\usepackage{makecell} 
\usepackage{booktabs}
\usepackage{multirow}
\usepackage[table]{xcolor}
\usepackage[symbol]{footmisc}

\setcopyright{none} 
\copyrightyear{2026}
\acmYear{2026}
\acmDOI{XXXXXXX.XXXXXXX}

\acmConference[MICRO 2026]{The 58th IEEE/ACM International Symposium on Microarchitecture}{October 31--November 04, 2026}{Athens, Greece}

\acmISBN{978-X-XXXX-XXXX-X/XX/XX}
\usepackage{tikzit}
\usepackage{tikzquanto}
\input{zx.tikzdefs}

\tikzstyle{box}=[shape=rectangle, text height=1.5ex, text depth=0.25ex, yshift=0.5mm, fill=white, draw=black, minimum height=5mm, yshift=-0.5mm, minimum width=5mm, font={\small}]
\tikzstyle{Z dot}=[inner sep=0mm, minimum size=2mm, shape=circle, draw=black, fill={rgb,255: red,221; green,255; blue,221}]
\tikzstyle{Z phase dot}=[minimum size=5mm, font={\footnotesize\boldmath}, shape=rectangle, rounded corners=2mm, inner sep=0.2mm, outer sep=-2mm, scale=0.8, tikzit shape=circle, draw=black, fill={rgb,255: red,221; green,255; blue,221}, tikzit draw=blue]
\tikzstyle{X dot}=[Z dot, shape=circle, draw=black, fill={rgb,255: red,255; green,136; blue,136}]
\tikzstyle{X phase dot}=[Z phase dot, tikzit shape=circle, tikzit draw=blue, fill={rgb,255: red,255; green,136; blue,136}, font={\footnotesize\boldmath}]
\tikzstyle{hadamard}=[fill=yellow, draw=black, shape=rectangle, inner sep=0.6mm, minimum height=1.5mm, minimum width=1.5mm]
\tikzstyle{vertex}=[inner sep=0mm, minimum size=1mm, shape=circle, draw=black, fill=black]
\tikzstyle{vertex set}=[inner sep=0mm, minimum size=1mm, shape=circle, draw=black, fill=white, font={\footnotesize\boldmath}]
\tikzstyle{Xexp}=[draw=black, shape=rectangle, minimum height=3mm, minimum width=3mm, fill={rgb,255: red,255; green,136; blue,136}]
\tikzstyle{Zexp}=[draw=black, shape=rectangle, minimum height=3mm, minimum width=3mm, fill={rgb,255: red,221; green,255; blue,221}]
\tikzstyle{Yexp}=[antipode, minimum height=3mm, minimum width=3mm, tikzit fill={rgb,255: red,100; green,205; blue,205}, tikzit shape=rectangle]
\tikzstyle{Aexp}=[vertex, fill=white, draw=black, shape=rectangle, rounded corners=1mm, minimum height=5mm, minimum width=5mm]

\tikzstyle{hadamard edge}=[-, dashed, dash pattern=on 2pt off 0.5pt, thick, draw={rgb,255: red,68; green,136; blue,255}]
\tikzstyle{brace edge}=[-, tikzit draw=blue, decorate, decoration={brace,amplitude=1mm,raise=-1mm}]
\tikzstyle{diredge}=[->]

\usetikzlibrary {arrows.meta}

\newcommand{\bra}[1]{\ensuremath{\left\langle #1 \right|}}
\newcommand{\ket}[1]{\ensuremath{\left|  #1 \right\rangle}}

\newcommand{\ketbra}[2]{\ensuremath{\ket{#1}\!\bra{#2}}}

\newcommand{\rev}[1]{\textcolor{black}{#1}}

\newcommand*\circled[1]{\tikz[baseline=(char.base)]{\node[shape=circle,draw,inner sep=0.75pt] (char) {#1};}}

\AtBeginDocument{%
  }

\begin{document}

\title{GadIR: A Spatial-Topology Preserving Compiler for Quantum Many-Body Systems Simulation}

\author{Xiangyu Ren\footnote[1]{}\textsuperscript{1, 7}, Yuexun Huang\textsuperscript{2}, Zhaohui Yang\textsuperscript{3}, Yuchen Zhu\textsuperscript{4}, \\ 
Tsung-Wei Huang\textsuperscript{5}, Tsung-Yi Ho\textsuperscript{6}, Zhiding Liang\footnote[1]{}\textsuperscript{6}, Antonio Barbalace\textsuperscript{1}}
\affiliation{
  \institution{\textsuperscript{1}The University of Edinburgh, Edinburgh, United Kingdom}
  \institution{\textsuperscript{2}The University of Chicago, Chicago, IL, USA}
  \institution{\textsuperscript{3}The Hong Kong University of Science and Technology, Kowloon, Hong Kong}
  \institution{\textsuperscript{4}Northwestern University, Chicago, IL, USA}
  \institution{\textsuperscript{5}University of Wisconsin–Madison, Madison, WI, USA}
  \institution{\textsuperscript{6}The Chinese University of Hong Kong, Shatin, Hong Kong}
  \institution{\textsuperscript{7}Open Quantum Intelligence Co., Ltd.}
  \country{}
}

\hyphenation{modu-larized}
\hyphenation{front-end}
\hyphenation{eva-luate}
\hyphenation{using}
\hyphenation{design}
\begin{abstract}
Simulating quantum many-body systems has been one of the most important applications of quantum computation. 
For simulation, the Hamiltonian of a physical system is compiled into quantum programs with native instructions for quantum hardware.
In previous works, the Hamiltonian is represented as Pauli strings, then compiled and optimized based on the quantum circuit model.
Such representation paradigm neglects the spatial topology of original physical models, which is vital information to reducing the overhead of compiling many-body systems Hamiltonians. 

To address such neglect, we introduce a spatial-topology preserving compiler for quantum many-body simulation. 
Using Pauli gadgets as the representations of the Hamiltonian, we introduce our intermediate representation -- GadIR, to preserve the spatial-topology information of original physical models.
Our compiler frontend performs the group reduction algorithm based on Pauli gadget model, which is a hardware-independent optimization.
Our compiler backend performs trotterization and scheduling on Pauli gadgets, then synthesizes the Pauli gadgets into hardware-native quantum programs.
We evaluate our compiler on all the canonical quantum many-body system models,
while achieving a significant reduction on compilation overhead regarding four major quantum architectures.
\rev{Overall, our spatial-topology preserving IR exploits the compilation optimization space for quantum many-body systems Hamiltonian.}

\end{abstract}

\maketitle

\begin{figure*}[t]
    \centering
    \includegraphics[width=0.7\textwidth]{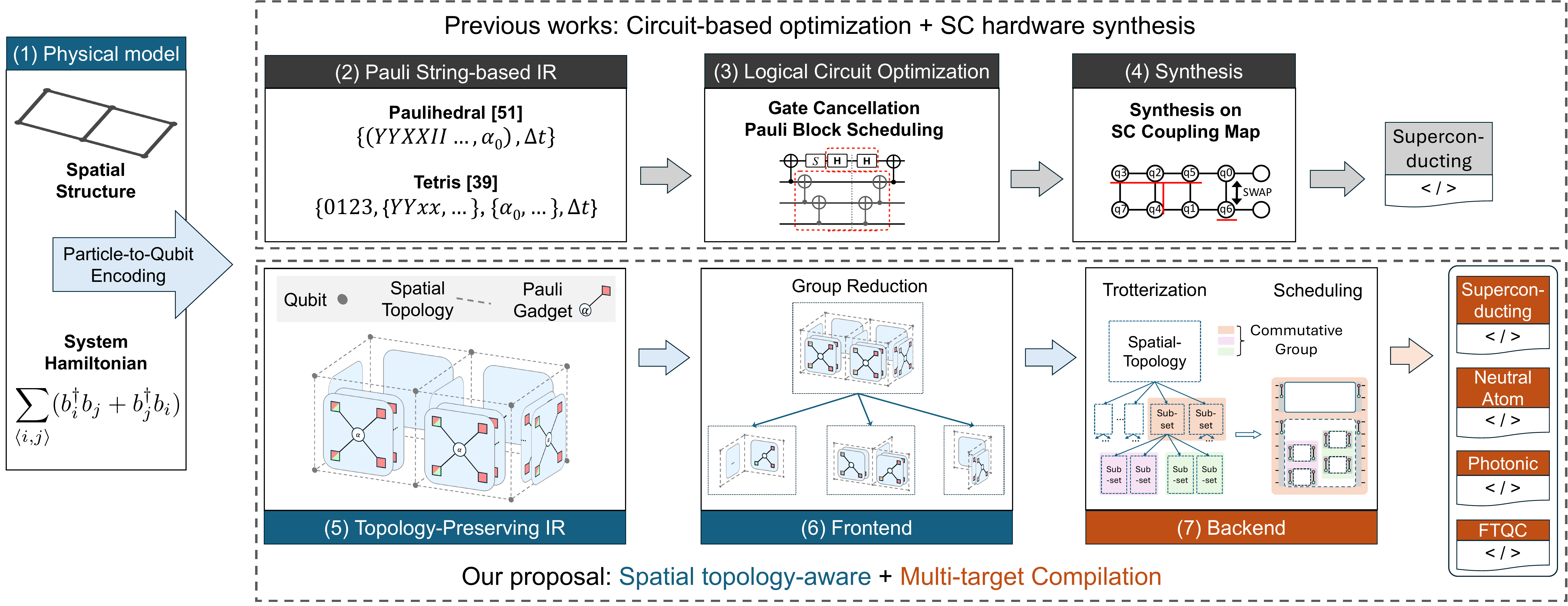}
    \caption{Comparison between our proposed framework and the frameworks in previous research~\cite{li2022paulihedral,lao20222qan,jin2024tetris,liu2025quclear}.}
    \label{fig:overview}
    \Description{Overview of the system.}
\end{figure*}

\hyphenation{simu-lating}
\hyphenation{covers}
\section{Introduction}  
Among the various applications of quantum computing, simulating physical systems has long stood out as a primary motivation~\cite{feynman2018simulating}.
Among various physical systems, the simulations of physical many-body systems are gaining prevalence, especially after their utility has been demonstrated on IBM quantum computers~\cite{kim2023evidence}.
The quantum many-body simulation formulates a physical system on specific spatial topology, such as an $n$-dimensional lattice. 
Taking the Bose-Hubbard model~\cite{elstner1999dynamics} as an example, the bosons are positioned at the sites of a 2D lattice, while they only interact with their neighboring bosons (box (1) of \Cref{fig:overview}).
This topology is applicable to physical models of bosons and fermions -- two fundamental classes of particles.
The core target of quantum many-body simulation is to simulate above physical models on quantum computers.

\textbf{Problem Definition.} In a physical model, the interactions between particles are formulated as Hamiltonian evolution, then the Hamiltonian can be approximated by quantum gates. Specifically, when a physical model is encoded as the Hamiltonian $H$, its evolution over time $t$ can be simulated by the operator $U=e^{-iHt}$. A compiler takes in the operator $U$ and compiles it into a hardware-native quantum program to be executed on the target hardware. In this work, our main problem is compiling the Hamiltonians of physical many-body models into hardware-native programs, for different quantum hardware architectures.

\textbf{Prior Works.} A series of research projects have been working on similar problems. Paulihedral~\cite{li2022paulihedral} and 2QAN~\cite{lao20222qan} leverage the permutation of Hamiltonian terms, to explore gate cancellation or SWAP reduction by reordering the Pauli strings.
Tetris~\cite{jin2024tetris} improves gate cancellation technique and reduces hardware mapping cost.
QuCLEAR~\cite{liu2025quclear} introduces a Clifford extraction method to reduce Pauli operators, based on the weak commutation of Clifford gates. 
Generally, these compilers use flattened Pauli strings as their intermediate representation (IR), optimizing mainly for circuit-level metrics such as two-qubit-gate count, circuit depth, or routing overhead.


\textbf{Challenge 1: Preserving topology of physical model.} 
As the Hamiltonian $H$ derives from a physical model with specific topology of particles, it contains spatial-topology information about the model structure.
Such information assists in reducing compilation overhead, which will be elaborated in~\Cref{sec:frontend}.
However, prior works neglect the spatial-topology information in the Hamiltonian when they represent it with Pauli strings. 
When represented as Pauli strings, the Hamiltonian $H$ is flattened into a linear sequence of Pauli operators, losing the spatial positional information of Pauli operators from original physical model.
This process is illustrated with examples in the box (2) of Figure~\ref{fig:overview}.

\textbf{Challenge 2: Hardware-independent optimization.}
Prior works reduce circuit compilation overhead -- mainly based on superconducting (SC) hardware settings~\cite{li2022paulihedral, lao20222qan, jin2024tetris, liu2025quclear}. 
This process is illustrated in the box (4) of Figure~\ref{fig:overview}.
Despite the efforts, those strategies they proposed are often incompatible with other quantum computing hardware, e.g., measurement-based quantum computation (MBQC), because it has a naturally different computation model.
Hence, a hardware-independent optimization that can be applied to diverse hardware architectures is highly desirable. 

\textbf{Solution: Spatial-topology preserving IR.}
To address the challenges mentioned above, we introduce a Pauli \ul{gad}get-based \ul{i}ntermediate \ul{r}epresentation -- GadIR, and implement the compiler based on GadIR.
The structure of GadIR is designed to \textbf{preserve the spatial topology of Hamiltonians derived from physical models}, while preserving such information throughout the compilation pipeline. 
The Pauli gadgets in GadIR have a graph structure corresponding to the physical model (shown in box (5) of Figure~\ref{fig:overview}), inspired by \textit{graph-based IR} in classical compiler techniques~\cite{click1995simple,bravcevac2023graph,leissa2015graph}.
The compiler frontend performs Pauli gadget group reduction on GadIR. 
The compiler backend performs hardware-independent trotterization and scheduling optimization, then synthesizes GadIR into hardware-native programs. 
Both frontend and backend contribute to reducing compilation overhead, with the following designs:

\textbf{Frontend.} 
GadIR represents the Hamiltonian operator as \textit{Pauli gadget}~\cite{cowtan2019phase}, which is a special graph structure of ZX-diagram~\cite{duncan2020graph}.
Specifically, compared to Pauli strings that linearly index the qubits, GadIR constructs a graph retaining the original topology of physical many-body models,
while Pauli gadgets are anchored on this topology.
Based on the topology information, our frontend optimizer performs\textit{ Pauli gadget reduction}, by eliminating certain patterns on the gadgets to reduce Hamiltonian operators. 
In contrast to previous works~\cite{li2022paulihedral, jin2024tetris} that apply gate cancellation to a pair of Pauli strings each time, our method can perform reduction on a large set of gadgets simultaneously, extending the optimization space.

\hyphenation{ins-tructions}
\textbf{Backend.} 
The backend performs Trotterization and scheduling directly on GadIR in ZX-diagram form, while the optimizations are all hardware-independent.
Then, the backend synthesizes GadIR into hardware-native programs, while the preserved topology information guides the compiler to find an appropriate qubit initial placement.
Compared to prior works that generate a gate-based logical circuit and synthesize it into a physical circuit for SC hardware, our backend supports multiple hardware architectures.

\hyphenation{hard-ware}
Although Pauli gadgets have been used to synthesize quantum chemistry circuits~\cite{cowtan2019phase,cowtan2020generic}, we are the first to use them for designing an IR and preserving model topology.
Our major contributions can be summarized as follows:
\begin{enumerate}
    \item For quantum many-body systems simulation, we propose a Pauli gadget-based IR to represent Hamiltonian terms, while preserving the spatial-topology information of original physical model.
    \rev{To the best of our knowledge, it is the first quantum compiler preserving Hamiltonian topology.}
    
    \item In compiler frontend, we introduce a novel optimization strategy, which is the Pauli gadget group reduction, based on our uniquely designed IR. 
    The strategy explores more chances for compilation overhead reduction than gate canceling~\cite{li2022paulihedral,jin2024tetris} and Clifford extraction~\cite{liu2025quclear}. 
    Furthermore, GadIR is hardware-agnostic and can be directly reused for multiple architectures, moving beyond conventional gate-based NISQ synthesis.
    
    \item In compiler backend, we perform hardware-independent scheduling using ZX-diagrams, reducing circuit depth. 
    Then we integrate hardware-specific optimizations in synthesis, utilizing preserved topology information.
\end{enumerate}

We evaluate our compiler framework on a wide range of benchmarks covering all canonical many-body models: spin, bosonic, fermionic and lattice-gauge. 
The evaluations are conducted on multiple hardware architectures, including SC architectures from IBM and Google devices, zoned-architecture neutral-atom, photonic MBQC on silicon quantum dots, and FTQC with lattice surgery scheme. 
The results show that our compiler achieves significant reduction of dominant compilation overheads for each architecture.
\ul{Compared to the state-of-the-art QuCLEAR, we reduce the overheads to 0.56$\times$ (\#2Q gate for SC), 0.87$\times$ (infidelity for neutral-atom), 0.85$\times$ (\#non-Clifford for FTQC), and 0.75$\times$ (\#emitter-entanglement for MBQC) of QuCLEAR on average.}
While our compiler is primarily designed for quantum many-body simulation, results show that it also works well on other quantum simulation problems, e.g., quantum chemistry. Furthermore, it is orthogonal to particle-to-qubit encoding techniques, e.g. Fermihedral~\cite{liu2024fermihedral} and HATT~\cite{liu2025hatt}.

\section{Background}

\subsection{Quantum Many-Body Simulation}\label{subsec:qmbs}
\hyphenation{topolo-gical}
\paragraph{Physical Many-Body Models.} 
Quantum simulation algorithms utilize digital quantum computers to simulate the particle interactions in physical models. 
In a physical model, these particles follow a specific spatial structure, e.g., residing on the sites of a lattice, and interact with their neighboring particles.
As a result, when we use Hamiltonians $H$ to describe the interactions, it will inherently contain topological information of the model structure. 
We refer to such information as the topology of original physical model.
An example of the spatial topology for a bosonic many-body system is shown in the box (1) of Figure~\ref{fig:overview}, as the shape of a $2\times3$ lattice.

\paragraph{Particle-to-Qubit Encoding.} 
In quantum mechanics, particles in nature are categorized into fermions and bosons. 
Hence, the Hamiltonians of a physical model are either fermionic operators or bosonic operators. 
To simulate these operators on quantum computers, they should be encoded to qubit operators, which are often represented in terms of Pauli operators.
This process is shown as the blue arrow connected to box (1) of Figure~\ref{fig:overview}.
The encoding methods include well-known Jordan-Wigner (JW), Bravyi-Kitaev (BK)~\cite{bravyi2002fermionic}, and recently proposed Fermihedral~\cite{liu2024fermihedral} and HATT~\cite{liu2025hatt}.

\paragraph{Inherent Topology of Qubit-Hamiltonian.}
After particle operators are encoded to qubit operators, the inherent topological information of Hamiltonian is also mapped to a new structure. 
For example, the Bravyi-Kitaev (BK) encoding~\cite{bravyi2002fermionic} maps fermionic model to a \textit{Fenwick tree} structure; bosonic linear encoding~\cite{somma2003bosoniclinear} maps bosonic model in an $n$-dimensional lattice to an $(n+1)$-dimensional lattice. 
In Figure~\ref{fig:overview}, the $2\times3$ bosonic model (depicted in box (1)) is mapped to a $2\times3\times2$ lattice of qubits (depicted in box (5)), which is the topology of its qubit-Hamiltonian.
These structures mentioned above are the spatial topologies of qubit-Hamiltonian, which is the topological information we preserve in GadIR.

\paragraph{Suzuki-Trotter Decomposition.}
To simulate the qubit-Hamiltonian $H$ on quantum hardware, it should be decomposed into unitary operations corresponding to each Hamiltonian term $H_i$.
We use \textit{Suzuki-Trotter} -- a foundational decomposition technique, to approximate the time evolution operator of qubit Hamiltonian. 
Given a Hamiltonian \( H = \sum_i H_i \), the time evolution operator \( e^{-iHt} \) can be approximated by a product of exponentials of the individual terms, with each term in the form like $H_i=Z_1Z_2X_3I_4Y_5$.
The second-order decomposition is given by
\begin{align}
    e^{-iHt} \approx \left( \prod_{i=1}^N e^{-iH_i t / (2r)} \prod_{i=N}^1 e^{-iH_i t / (2r)} \right)^r,
\end{align}
where \( r \) is the number of Trotter steps. 
In our compiler, we perform the trotterization in backend, as shown in the box (7) of Figure~\ref{fig:overview}.

\subsection{ZX-Diagram and Pauli Gadget}
In this subsection, we give a brief introduction to ZX-diagram, and Pauli gadget -- a special structure in the ZX-diagram. Further details of these concepts can be found in \cite{duncan2020graph,van2020zx,kissinger2019pyzx}.

\paragraph{ZX-Diagram}
The ZX-calculus is a diagrammatic language similar to the quantum circuit notation, while its instances are ZX-diagrams. 
The basic elements of a ZX-diagram are \textit{spiders} and \textit{wires}.
\textit{Spiders} are linear maps which can have any number of input or output \textit{wires}, and being associated with coefficients $\alpha$. 
There are two varieties of spiders -- the Z-spider depicted in green dot (\Cref{zsp}) and the X-spider depicted in red dot (\Cref{xsp}), while the definitions of their input-output maps are given on the right sides of equations.
\begin{equation}\label{zsp}
\textrm{\small
$\begin{tikzpicture}
	\begin{pgfonlayer}{nodelayer}
		\node [style=Z phase dot] (0) at (0, 0) {$\alpha$};
		\node [style=none] (1) at (1.25, 1) {};
		\node [style=none] (2) at (-1.25, 1) {};
		\node [style=none] (3) at (-1.25, -1) {};
		\node [style=none] (4) at (1.25, -1) {};
		\node [style=none] (5) at (1.25, 0.5) {};
		\node [style=none] (6) at (-1.25, 0.5) {};
		\node [style=none, rotate=90] (7) at (-1, -0.25) {...};
		\node [style=none, rotate=90] (8) at (1, -0.25) {...};
	\end{pgfonlayer}
	\begin{pgfonlayer}{edgelayer}
		\draw [in=-141, out=0, looseness=0.75] (3.center) to (0);
		\draw [in=180, out=-39, looseness=0.75] (0) to (4.center);
		\draw [in=180, out=22, looseness=0.75] (0) to (5.center);
		\draw [in=180, out=39, looseness=0.75] (0) to (1.center);
		\draw [in=0, out=158, looseness=0.75] (0) to (6.center);
		\draw [in=141, out=0, looseness=0.75] (2.center) to (0);
	\end{pgfonlayer}
\end{tikzpicture} \ := \ \ketbra{\textrm{$0$...$0$}}{\textrm{$0$...$0$}} +
e^{i \alpha} \ketbra{\textrm{$1$...$1$}}{\textrm{$1$...$1$}}$
}
\end{equation}

\begin{equation}\label{xsp}
\textrm{\small
$\begin{tikzpicture}
	\begin{pgfonlayer}{nodelayer}
		\node [style=X phase dot] (0) at (0, 0) {$\alpha$};
		\node [style=none] (1) at (1.25, 1) {};
		\node [style=none] (2) at (-1.25, 1) {};
		\node [style=none] (3) at (-1.25, -1) {};
		\node [style=none] (4) at (1.25, -1) {};
		\node [style=none] (5) at (1.25, 0.5) {};
		\node [style=none] (6) at (-1.25, 0.5) {};
		\node [style=none, rotate=90] (7) at (-1, -0.25) {...};
		\node [style=none, rotate=90] (8) at (1, -0.25) {...};
	\end{pgfonlayer}
	\begin{pgfonlayer}{edgelayer}
		\draw [in=-141, out=0, looseness=0.75] (3.center) to (0);
		\draw [in=180, out=-39, looseness=0.75] (0) to (4.center);
		\draw [in=180, out=22, looseness=0.75] (0) to (5.center);
		\draw [in=180, out=39, looseness=0.75] (0) to (1.center);
		\draw [in=0, out=158, looseness=0.75] (0) to (6.center);
		\draw [in=141, out=0, looseness=0.75] (2.center) to (0);
	\end{pgfonlayer}
\end{tikzpicture} \ := \ \ketbra{\textrm{$+$...$+$}}{\textrm{$+$...$+$}} +
e^{i \alpha} \ketbra{\textrm{$-$...$-$}}{\textrm{$-$...$-$}}$}
\end{equation}

A ZX-diagram can be built from spiders and wires, by composition of the elements shown above. 
For a quantum program with $m$ input qubits and $n$ output qubits, its corresponding ZX-diagram has $m$ input wires and $n$ output wires, representing linear maps $(\mathbb C^2)^{\otimes m} \to (\mathbb C^2)^{\otimes n}$.
As examples, we have these quantum gates represented in ZX-diagram:
$$
    \begin{array}{cclcl}
    \begin{tikzpicture}
	\begin{pgfonlayer}{nodelayer}
		\node [style=Z phase dot] (0) at (0, 0) {$\alpha$};
		\node [style=none] (1) at (-1.25, 0) {};
		\node [style=none] (2) at (1.25, 0) {};
	\end{pgfonlayer}
	\begin{pgfonlayer}{edgelayer}
		\draw (1.center) to (0);
		\draw (0) to (2.center);
	\end{pgfonlayer}
\end{tikzpicture} & = & \ketbra{0}{0} + e^{i \alpha} \ketbra{1}{1} & = & R_Z(\alpha) \\
    \\
    \begin{tikzpicture}
	\begin{pgfonlayer}{nodelayer}
		\node [style=X phase dot] (0) at (0, 0) {$\alpha$};
		\node [style=none] (1) at (-1, 0) {};
		\node [style=none] (2) at (1, 0) {};
	\end{pgfonlayer}
	\begin{pgfonlayer}{edgelayer}
		\draw (1.center) to (0);
		\draw (0) to (2.center);
	\end{pgfonlayer}
\end{tikzpicture}
 & = & \ketbra{+}{+} + e^{i \alpha} \ketbra{-}{-} & = & R_X(\alpha)
    \end{array}
$$



Furthermore, to represent a CNOT gate, we can 
connect the second output wire of Z-spider to the first input wire of X-spider:

\begin{equation}\label{eq:cnot}
    \begin{tikzpicture}
	\begin{pgfonlayer}{nodelayer}
		\node [style=Z dot] (0) at (0, 0.5) {};
		\node [style=none] (1) at (1.25, 0.5) {};
		\node [style=X dot] (2) at (0, -0.5) {};
		\node [style=none] (3) at (-1.25, 0.5) {};
		\node [style=none] (4) at (-1.25, -0.5) {};
		\node [style=none] (5) at (1.25, -0.5) {};
	\end{pgfonlayer}
	\begin{pgfonlayer}{edgelayer}
		\draw (0) to (2);
		\draw (0) to (3.center);
		\draw (0) to (1.center);
		\draw (5.center) to (2);
		\draw (2) to (4.center);
	\end{pgfonlayer}
\end{tikzpicture} \quad =\quad 
    CNOT \quad 
\end{equation}


\paragraph{Pauli Gadget}
Similar to Pauli string and Pauli kernel~\cite{li2022paulihedral} for quantum circuit, the Pauli gadget is a high-level structure for ZX-diagram, representing a Hamiltonian term.
Given a Hamiltonian term $e^{-i\alpha X_0Y_1I_2Z_3}$, it can be depicted as:
\begin{equation}
    e^{\frac{-i\alpha}{2} X_0Y_1I_2Z_3} \quad = \quad \vcenter{\hbox{\scalebox{0.85}{\def\scl{0.7}
\begin{tikzpicture}[baseline={(2)}]
	\begin{pgfonlayer}{nodelayer}
        \node [style=none] (14) at (-0.5, 0*\scl) {$q_3$};
        \node [style=none] (15) at (-0.5, 1.5*\scl) {$q_2$};
        \node [style=none] (16) at (-0.5, 3*\scl) {$q_1$};
        \node [style=none] (17) at (-0.5, 4.5*\scl) {$q_0$};
		\node [style=none] (0) at (0, 0 *\scl) {};
		\node [style=none] (1) at (0, 1.5 *\scl) {};
		\node [style=none] (2) at (0, 3 *\scl) {};
		\node [style=none] (3) at (0, 4.5 *\scl) {};
		\node [style=Zexp] (4) at (1.5, 0 *\scl) {};
		\node [style=Xexp] (5) at (1.5, 4.5 *\scl) {};
		\node [style=Yexp] (6) at (1.5, 3 *\scl) {};
		\node [style=none] (9) at (5, 4.5 *\scl) {};
		\node [style=none] (10) at (5, 3 *\scl) {};
		\node [style=none] (11) at (5, 1.5 *\scl) {};
		\node [style=none] (12) at (5, 0 *\scl) {};
		\node [style=Aexp] (13) at (3.5, 2 *\scl) {$\alpha$};
	\end{pgfonlayer}
	\begin{pgfonlayer}{edgelayer}
		\draw (3.center) to (9.center);
		\draw (2.center) to (10.center);
		\draw (1.center) to (11.center);
		\draw (12.center) to (4);
		\draw (4) to (0.center);
		\draw (5) to (13);
		\draw (13) to (6);
		\draw (4) to (13);
	\end{pgfonlayer}
\end{tikzpicture}}}} \quad,
\end{equation}
where the spiders of Pauli gadget are especially shaped in squares. While the green square and red square still correspond to Z-spider and X-spider, the Y-spider is colored in both green and red. 
Each of the Pauli operators $\sigma_X, \sigma_Y, \sigma_Z$ is represented as X-, Y- and Z- spiders respectively. In the meantime, they are connected to the box of $\alpha$, representing the coefficient of the Hamiltonian term.


\begin{figure*}
\centering
    \includegraphics[width=0.9\textwidth]{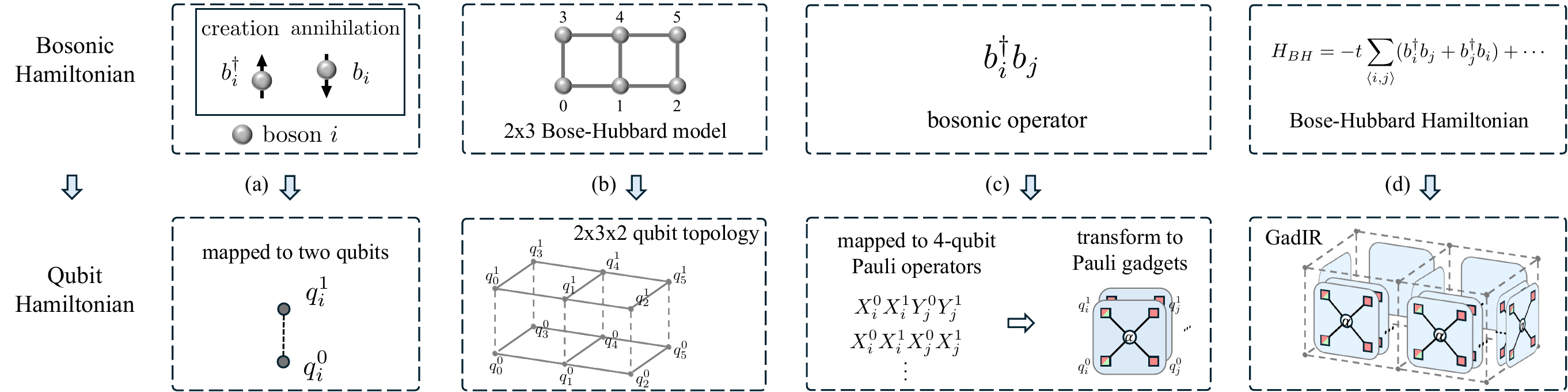}
    \caption{Mapping a $2\times3$ Bose-Hubbard model to qubit Hamiltonian, and construct the corresponding GadIR.}
    \label{fig:bh_mapping}    
\end{figure*}

\section{GadIR Design}\label{sec:ir}
In this section, we describe the design and motivation of our topology-preserving GadIR.
Using the example of a $2\times3$ bosonic model, we introduce how the original physical model and Hamiltonian terms are converted to GadIR step-by-step.
\rev{
The GadIR syntax is introduced in \ref{subsec:ir_syntax}.
Then in \ref{subsec:topo_motivation} we demonstrate the insight of GadIR: it captures the optimizations that Pauli-string IR misses.
Following is the GadIR generation for physical models in \ref{subsec:ir_gen}.
}

\subsection{Introducing Topology-Preserving IR} \label{subsec:ir_design}
GadIR consists of two parts: \circled{1} spatial topology as the "frame", derived from the physical structure, depicted in the lower part of~\Cref{fig:bh_mapping}b; \circled{2} Pauli gadgets as the "body", constructed from Hamiltonian terms, depicted in~\Cref{fig:bh_mapping}c as the blue areas.
While the spatial topology provides supportive information of inherent structures within the model, the Pauli gadget provides a concise representation of each Hamiltonian term.
These two parts combine to form GadIR, shown in \Cref{fig:bh_mapping}d.

To understand how the topology frame is built, first we introduce the physical structure of a many-body system. 
We select the Bose-Hubbard model~\cite{elstner1999dynamics} as an example, which describes the interaction between boson particles.
In the Bose-Hubbard model, the bosons are located on the sites of a lattice, and in this example we have a $2 \times 3$ lattice as the physical structure.
The Hamiltonian of Bose-Hubbard model is described as
\begin{align}\label{eq:bh}
H_{BH} = - t \sum_{\langle i, j \rangle} (b_i^\dagger b_j + b_j^\dagger b_i)
 + \frac{U}{2} \sum_i b_i^\dagger b_i (b_i^\dagger b_i - 1)
 - \mu \sum_i b_i^\dagger b_i.
\end{align}

Here in the $H_{BH}$, $\langle i,j \rangle$ denotes a pair of neighboring sites of the lattice.
For the site $i$, we use $b_i^\dagger$ to denote the bosonic creation, and $b_i$ to denote bosonic annihilation at this site.
From the Hamiltonian terms in~\Cref{eq:bh}, we can notice that bosonic operators are only associated with neighboring bosons in an individual term, implying the locality of the model is critical information.

\paragraph{Spatial Topology}
The above bosonic Hamiltonian of Bose-Hubbard model is transformed to qubit Hamiltonian, via a boson-to-qubit mapping method~\cite{somma2003bosoniclinear, miessen2021quantum}.
In this mapping, each boson, along with its creation and annihilation operators ($b_i^\dagger,b_i$), is mapped to two qubits ($q_i^0,q_i^1$) in the qubit Hamiltonian model (\Cref{fig:bh_mapping}a). 
Hence, the $2 \times 3$ bosonic Hamiltonian model is mapped to a $2 \times 3 \times 2$ qubit Hamiltonian model (\Cref{fig:bh_mapping}b). 
\ul{This $2 \times 3 \times 2$ lattice of qubits is exactly the spatial topology information included in GadIR}.

\paragraph{Pauli Gadget}
For the bosonic Hamiltonian terms in~\Cref{eq:bh}, such as $b_i^\dagger b_j$, are each mapped to a set of 4-qubit Pauli operators, such as $X_i^0X_i^1Y_j^0Y_j^1$ (\Cref{fig:bh_mapping}c). 
These Pauli operators are performed on $q_i^0,q_i^1,q_j^0,q_j^1$, which are the qubits derived from boson $i$ and $j$.
The terms in the second line of~\Cref{eq:bh}, such as $b_i^\dagger b_i$, are mapped to a set of 2-qubit Pauli operators performed on $q_i^0,q_i^1$.
Consequently, we transform these Pauli operators into Pauli gadgets, as illustrated by the blue areas in~\Cref{fig:bh_mapping}c. 
Such Pauli gadget representations are easily integrated into the frame of spatial topology, while their ZX-spiders can be attached to the qubit they perform on.

In~\Cref{fig:bh_mapping}d, when all the Pauli gadgets of qubit Hamiltonian are attached to the topology frame, we obtain the complete IR of this Bose-Hubbard model.
Compared to conventional Pauli string-based IR~\cite{li2022paulihedral,lao20222qan,jin2024tetris}, \textbf{GadIR can reflect the locality of Hamiltonian operators,} paving the way for the following optimizations in compiler frontend and backend.

\subsection{GadIR Syntax}\label{subsec:ir_syntax}
\begin{figure}[h]
    \centering
    \includegraphics[width=0.47\textwidth]{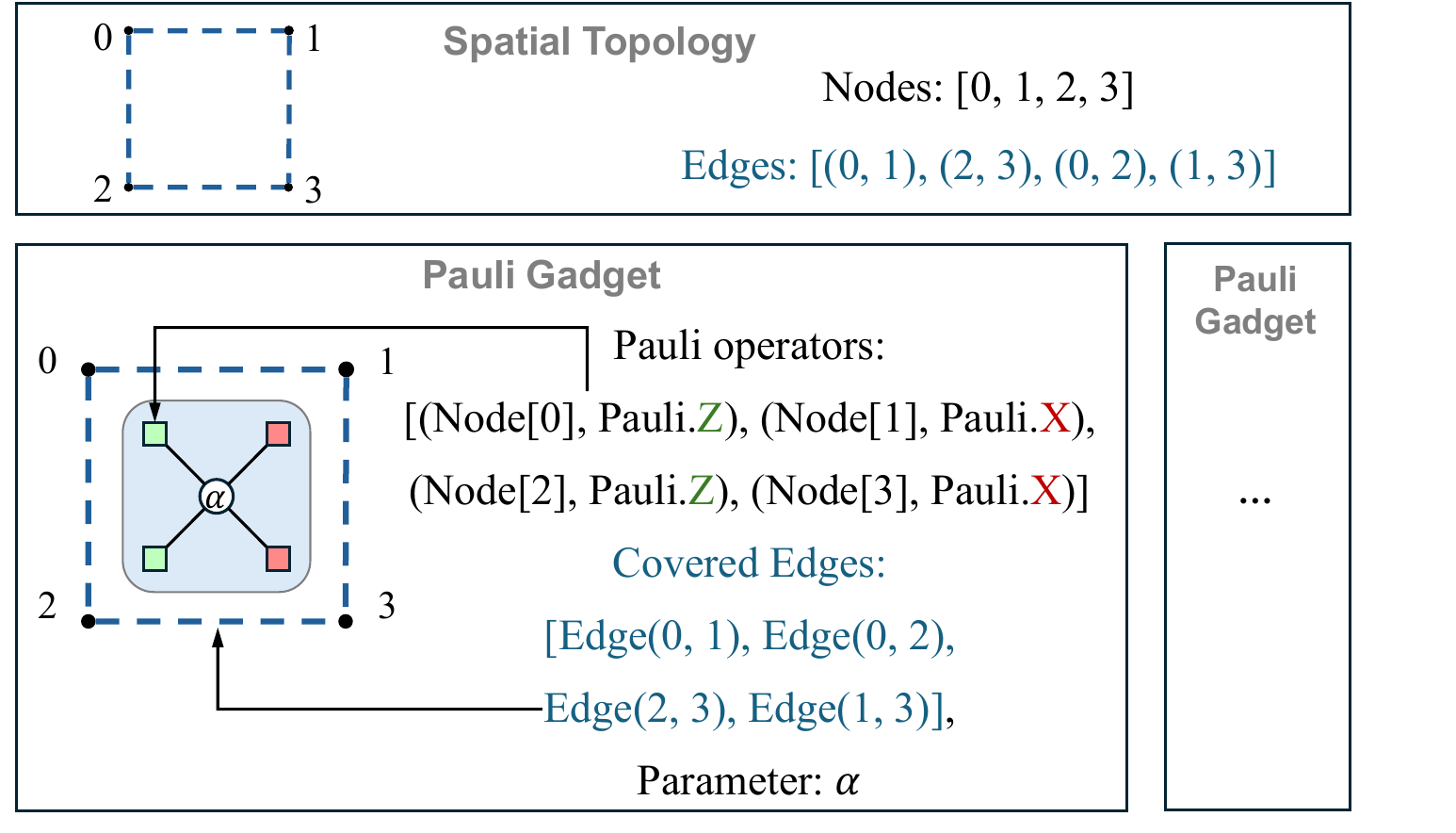}
    \caption{GadIR syntax. Up: the spatial topology graph for Hamiltonian of many-body system model. Down: the Pauli gadgets corresponding to Hamiltonian terms, including their operators and the edges they cover on topology graph.}
    \label{fig:ir_syntax}
\end{figure}

We designed a concise IR syntax to represent the spatial topology and Pauli gadgets, adopting the data structures shown in~\Cref{fig:ir_syntax}.
The topology is an undirected graph with its nodes as the qubits.
The Pauli gadgets contain two parts of information: 
(1) Pauli operators (X, Y, Z) that compose the Pauli gadget, shown as the spiders in ZX-diagram. 
(2) The edges in spatial topology that are covered by this Pauli gadget. Specifically, an edge begin covered by a Pauli gadget is defined as follows: both adjacent nodes of the edge have a spider (Pauli operator) from this Pauli gadget.
These covered edges serve as the critical information for group reduction method in \Cref{sec:frontend}.
\ul{We emphasize that storing the information of covered edges is the main motivation for choosing Pauli gadget as the IR representation, instead of choosing the Pauli string.}

\begin{figure}[h]
    \centering
    {\includegraphics[width=0.45\textwidth]{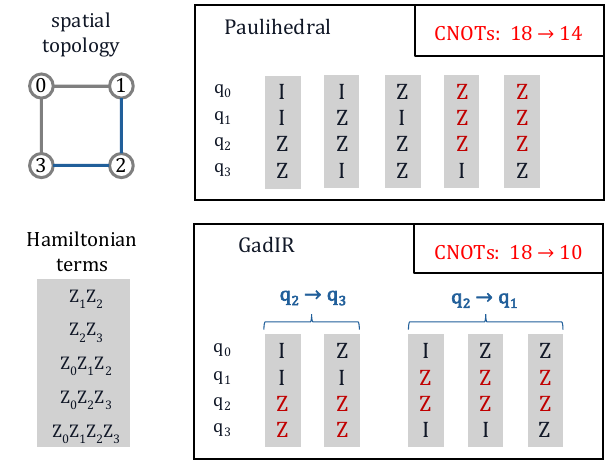}}
    \caption{\rev{Comparison of Paulihedral and GadIR on the ability of capturing gate cancellation opportunities. On the left are the Hamiltonian and its spatial topology. On the right are the sequences of Pauli terms ordered by Paulihedral or GadIR approach, while the gate cancellations are highlighted in red. The results show that GadIR gains more CNOT reductions.}}
    \label{fig:topo_motivation}
\end{figure}

\rev{\subsection{A Motivating Example}\label{subsec:topo_motivation}} 
\rev{The preserved topology information in GadIR helps compilers to explore more chances of optimization.
We explain the motivation of preserving spatial topology using an intuitive example shown in \Cref{fig:topo_motivation}. 
In this case, we apply the Paulihedral and GadIR methods to capture the opportunities of gate cancellation in a group of Hamiltonian terms. 
Note that here the Paulihedral method directly runs source code in ~\cite{li2022paulihedral}, while the GadIR method runs the algorithm we describe in \Cref{algo_ggr}. 
Based on the topological path $q_2 \rightarrow q_1$ and $q_2 \rightarrow q_3$, GadIR captures more gate cancellation chances than Paulihedral, gaining more CNOT reductions. 
The example shows that topology-preserving strategy outperforms Paulihedral's pairwise ordering strategy on general gate canceling optimization.
Furthermore, with the group reduction method designed for GadIR (\Cref{sec:frontend}), the topology information can be fully utilized.
}

\subsection{Generating IR from Physical Models}\label{subsec:ir_gen}
\begin{figure}[h]
    \centering
    \includegraphics[width=0.47\textwidth]{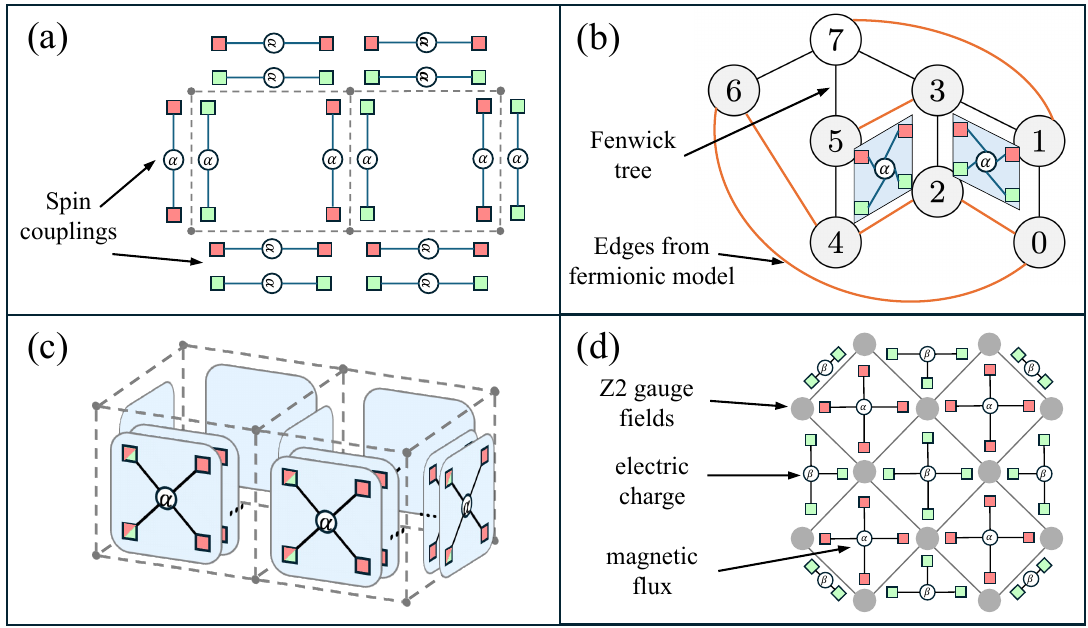}
    \caption{Examples of GadIR for (a) spin-qubit, (b) fermionic, (c) bosonic and (d) lattice gauge theory models.}
    \label{fig:ir_examples}
\end{figure}

We target all canonical models in quantum many-body physics~\cite{georgescu2014quantum,cirac2012goals}, with other specialized models often reducible to or expressible within them.
These models are illustrated in \Cref{fig:ir_examples}.

1. The \textbf{spin-qubit models}~\cite{porras2004effective} (e.g. \textit{Heisenberg, Ising}) define their physical interactions directly on qubit Hamiltonian, so the topology information is identical to the physical topology of spin-qubits. The Pauli operators are derived from spin couplings.

2. The \textbf{fermionic models}~\cite{troyer2005computational} define their interactions on local fermionic modes, and we utilize \textit{Bravyi-Kitaev} transformation~\cite{bravyi2002fermionic} to map them into a \textit{Fenwick tree}~\cite{fenwick1994new} structure of qubits.
For a fermion mode $i$, its creation operator $f_i^\dagger$ and annihilation operator $f_i$ are generally mapped to a subgraph of the Fenwick tree~\cite{havlivcek2017operator}.
We merge the set of edges in the Fenwick tree $E_{tree}$ (shown as black lines in \Cref{fig:ir_examples}b) with the set of edges in original fermion model $E_{model}$ (partly shown as orange lines in \Cref{fig:ir_examples}b), resulting in $E_{tree} \cup E_{model}$ as the spatial topology in the IR.

3. The \textbf{bosonic models}~\cite{guo2012critical} are typically located on the sites of a lattice, and are mapped to Hamiltonian terms via linear method~\cite{miessen2021quantum}.
We use the mapping method described in the example of~\Cref{subsec:ir_design}.

4. The \textbf{lattice gauge theory models}~\cite{cochran2025visualizing} arrange the \textit{matter fields} on the sites of a $x \times y$ lattice~\cite{cochran2025visualizing}. 
As shown in \Cref{fig:ir_examples}d, their interactions, namely $\mathbb{Z}_2$ \textit{gauge fields}, are mapped to qubits, which is itself a centered square lattice (gray nodes and edges). 
The \textit{electric charge} or \textit{magnetic flux} interactions are transformed into Pauli gadgets acting on squares in the topology (Pauli gadgets).

\section{Compiler Frontend Design} \label{sec:frontend}
\subsection{Preliminaries: Gadget Reduction}
In this section, we introduce our method to optimize the Pauli gadgets in GadIR, thus reducing the Hamiltonian terms.
We start with the Pauli gadget reduction~\cite{cowtan2019phase}.
When specific gates pass through two spiders of a Pauli gadget, and these two spiders are in specific patterns, one of the spiders will be removed:
\begin{equation}\label{cnotrulezz}
    \scalebox{0.8}{\begin{tikzpicture}[baseline={(1)}]
    \draw[-{Latex}]        (0,1.5)   -- (1,1.5);
	\begin{pgfonlayer}{nodelayer} 
		\node [style=none] (0) at (0, 1) {};
		\node [style=none] (1) at (0, 0) {};
		\node [style=X dot] (2) at (1, 0) {};
		\node [style=Z dot] (3) at (1, 1) {};
		\node [style=none] (7) at (2.5, -2) {};
		\node [style=none] (8) at (4.5, 1) {};
		\node [style=none] (9) at (4.5, 0) {};
		\node [style=none] (10) at (2.5, -1) {$\vdots$};
		\node [style=Zexp] (11) at (2, 0) {};
		\node [style=Aexp] (12) at (3.5, -1) {$\alpha$};
        \node [style=Zexp] (13) at (2, 1) {};
        \node [style=none] (14) at (-0.5, 1) {$0$};
        \node [style=none] (15) at (-0.5, 0) {$1$};
	\end{pgfonlayer}
	\begin{pgfonlayer}{edgelayer}
		\draw (0.center) to (3);
		\draw (3) to (8.center);
		\draw (3) to (2);
		\draw (2) to (1.center);
		\draw (2) to (11);
		\draw (11) to (9.center);
		\draw (11) to (12);
            \draw (13) to (12);
		\draw (12) to (7.center);
	\end{pgfonlayer}
\end{tikzpicture} \qquad = \qquad \begin{tikzpicture}[baseline={(1)}]
	\begin{pgfonlayer}{nodelayer}
		\node [style=none] (0) at (0, 1) {};
		\node [style=none] (1) at (0, 0) {};
		\node [style=X dot] (2) at (3.5, 0) {};
		\node [style=Z dot] (3) at (3.5, 1) {};
		\node [style=none] (7) at (1.5, -2) {};
		\node [style=none] (8) at (4.5, 1) {};
		\node [style=none] (9) at (4.5, 0) {};
		\node [style=none] (10) at (1.5, -1) {$\vdots$};
		\node [style=Zexp] (11) at (1, 0) {};
		\node [style=Aexp] (13) at (2.5, -1) {$\alpha$};
        \node [style=none] (14) at (-0.5, 1) {$0$};
        \node [style=none] (15) at (-0.5, 0) {$1$};
	\end{pgfonlayer}
	\begin{pgfonlayer}{edgelayer}
		\draw (3) to (2);
		\draw (0.center) to (3);
		\draw (3) to (8.center);
		\draw (9.center) to (2);
		\draw (2) to (11);
		\draw (11) to (1.center);
		\draw (13) to (11);
		\draw (13) to (7.center);
	\end{pgfonlayer}
\end{tikzpicture}}
\end{equation}
In~\Cref{cnotrulezz}, when a CNOT passes the spiders with Pauli operators $Z_0Z_1$, one of the spiders is eliminated, turning into operators $I_0Z_1$.
From a mathematical aspect, this is equivalent as transforming $[\text{CNOT}_{0,1} \cdot Z_0Z_1]$ to $[I_0Z_1 \cdot \text{CNOT}_{0,1}]$.
In this transforming, the Pauli operator $Z_0$ is reduced.
By replacing the CNOT here with other gate sets, we can reduce different Pauli operators.

\subsection{Gadget Group Reduction}\label{subsec:ir_frontend}
\begin{figure}[tbp]
    \centering
    \includegraphics[width=0.4\textwidth]{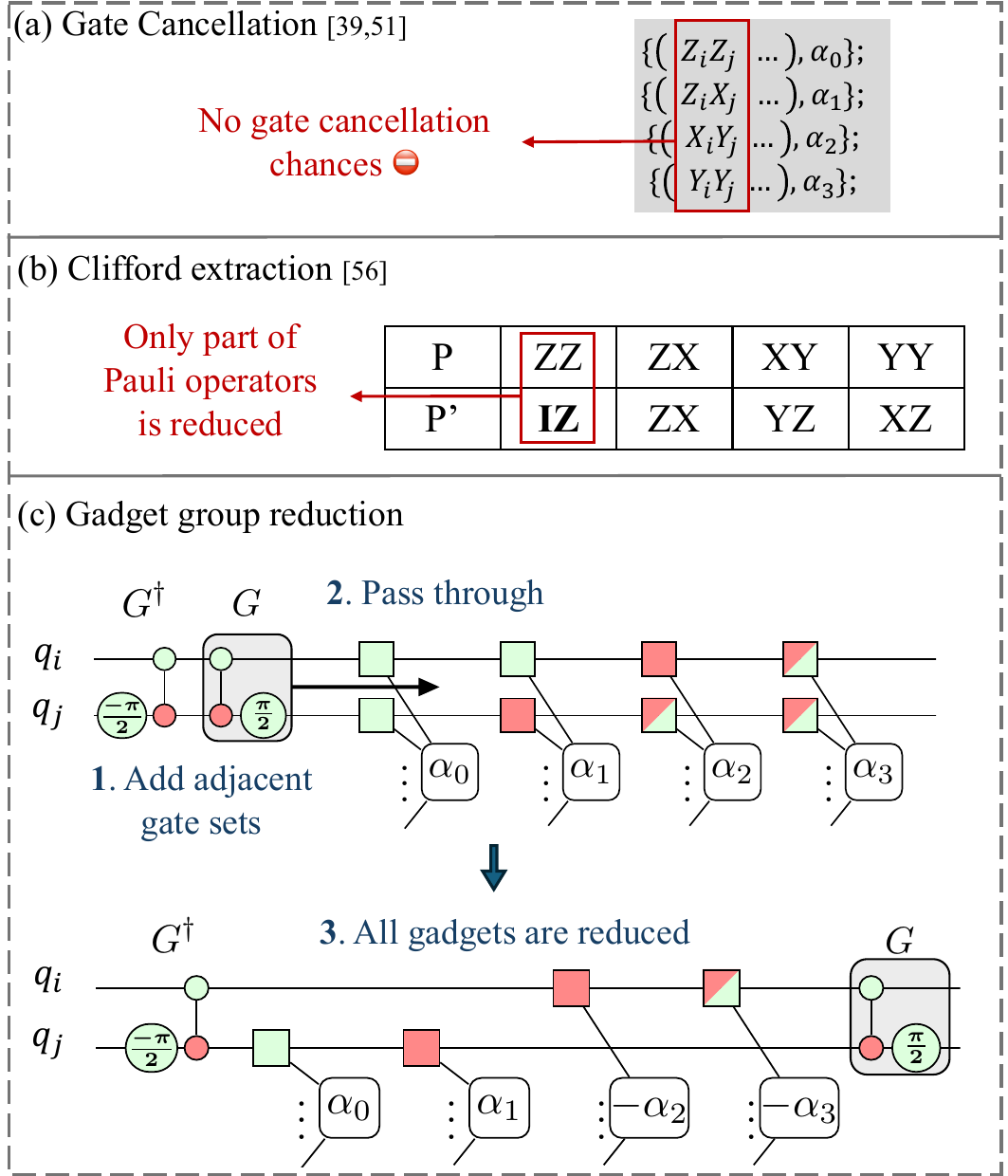}
    \caption{(c) Gadget reduction compared with (a) gate cancellation~\cite{li2022paulihedral, jin2024tetris} and (b) Clifford extraction~\cite{liu2025quclear}.}
    \label{fig:group_reduction}
\end{figure}

In \Cref{sec:ir} we analyze the locality of Hamiltonian in quantum many-body systems, so we can utilize the locality characterization to maximally reduce Pauli gadgets.
In our compiler frontend, we introduce an optimization method based on GadIR, which is the \textit{gadget group reduction}. 
When a gate set passes through multiple pairs of specific spiders pattern (from multiple gadgets) sequentially, one spider in each pair gets eliminated.
We provide an example in~\Cref{fig:group_reduction}c, where a group of Pauli gadgets having $[ZZ,ZX,XY,YY]$ spiders on qubit $q_i$ and $q_j$.
When we pass a gate set $G=[\text{CNOT}_{i,j}\cdot\text{R}_Z(\frac{\pi}{2})_j]$ through these spiders, one of the spiders is removed in each gadget, turning into $[IZ,IX,XI,YI]$.

Here we explain the details of gadget group reduction step by step, illustrated in~\Cref{fig:group_reduction}c. 
In step\textbf{1}, we add two adjoint gate sets $G=[\text{CNOT}_{i,j}\cdot\text{R}_Z(\frac{\pi}{2})_j]$ and $[G^\dagger=\text{R}_Z(\frac{-\pi}{2})_j\cdot\text{CNOT}_{i,j}]$ on the left of gadgets group.
Since $G^\dagger \cdot G=I$, adding them does not change the circuit logic.
Then in step\textbf{2}, we pass $G$ from left to right, through these gadgets.
As a result in step\textbf{3}, these gadgets are reduced based on gate set passing rules, and the two gate sets $G^\dagger$ and $G$ end up on each side of gadgets group.
Generally, removing one spider in the gadget corresponds to reducing one Pauli operator from the Hamiltonian term, equivalent to reducing two CNOTs in compiled circuit~\cite{li2022paulihedral}.
Hence, \ul{applying reduction to a group of $n$ gadgets can reduce $2n-2$ CNOTs at a time}, since two extra CNOTs are induced from $G^\dagger$ and $G$. 

\paragraph{A Full Set of Reduction Rules.}\label{subsec:reductionrules}
We derive the group reduction rules of Pauli gadgets from the commutation rules of Pauli gadgets~\cite{cowtan2019phase}.
In the last paragraph we show passing a gate set $G$ through multiple gadgets to remove their spiders and reduce the gadgets.
Specifically, when we \ul{customize the gate sets $G$ as different combinations of the gates} in $\{\text{CNOT},\text{H},\text{R}_X(\frac{\pi}{2}), \text{R}_Z(\frac{\pi}{2})\}$, it \ul{reduces different patterns of Pauli gadgets}.
We specify the correspondence between gate set $G$ and reducible gadgets in~\Cref{tab:reduction_rule}. 
\rev{The reduction rules are exhaustive under a Clifford model: We enumerate all \(24^2\) local-Clifford dressings of both CNOT directions, Pauli signs, and qubit-order reversal. The model yields exactly the six canonical reducible Pauli-pair classes, as shown in \Cref{tab:reduction_rule}.}


\paragraph{Comparison with Prior Methods}
Shown in \Cref{fig:group_reduction}a, the \textit{gate cancellation} techniques (e.g. Paulihedral~\cite{li2022paulihedral} and Tetris~\cite{jin2024tetris}) only cancel the CNOTs from identical parts of Pauli strings. In contrast, our group reduction eliminates different patterns of Pauli operators simultaneously. 
Shown in \Cref{fig:group_reduction}b, \textit{Clifford extraction} method in QuCLEAR~\cite{liu2025quclear} can only reduce a limited range of Pauli operator patterns (\Cref{fig:group_reduction}b).
Specifically, \textit{Clifford extraction} achieves the same reduction result as the first row of~\Cref{tab:reduction_rule} (colored in gray), but neglects other reducible patterns.
In contrast, our reduction method covers \textbf{all combinations of Pauli operators}. 
Previous works~\cite{cowtan2019phase,cowtan2020generic} apply isolated \textit{gadget reduction} on Variational Quantum Eigensolver (VQE) circuits, but their method fails to preserve the spatial-topology of Hamiltonian, thus losing chances of dedicated reductions on groups of gadgets.
Also, the naturally weak-locality characterization of VQE leads to limited reduction.

\paragraph{Insight with Hamiltonian Locality.} 
The insight for \textit{gadget group reduction} derives from the locality of Hamiltonian in quantum many-body systems. 
As we observed from \Cref{fig:bh_mapping}, an edge in the spatial topology graph is covered by multiple Pauli gadgets, providing an ideal opportunity to reduce these gadgets simultaneously using the gadget group reduction.
\ul{When the Hamiltonian locality of a physical model is stronger, the group of reducible Pauli gadgets is larger, thus the optimization will be more effective.}
\rev{In \Cref{sec:locality_analysis}, we give definitions about Hamiltonian locality, and analyze its impact on GadIR performance.}

\begin{table}[tbp]
\centering
\caption{Reduction rules for Pauli gadgets.}\label{tab:reduction_rule}
\bgroup
\def\arraystretch{1.2}
\begin{NiceTabular}{ |c|c| } 
 \hline \rowcolor{White}
 Gate Set $G$    & Reducible Gadget Patterns \\
 to Pass Through & ($-$) = Coefficient Negation \\ \hline 
 \rowcolor{LightGray}
 $CNOT_{i,j}$ & $Z_iZ_j,\  Z_iY_j,\  X_iX_j,\  Y_iX_j$ \\ \hline
 $H_i \cdot CNOT_{i,j}$ & $X_iZ_j,\  X_iY_j,\  Z_iX_j,\  Y_iX_j(-)$ \\ \hline
 $H_j \cdot CNOT_{i,j}$ & $Z_iX_j,\  Z_iY_j(-),\  X_iZ_j,\  Y_iZ_j$ \\ \hline
 $R_Z(\frac{\pi}{2})_j \cdot CNOT_{i,j}$ & $Z_iZ_j,\  Z_iX_j,\  X_iY_j,\  Y_iY_j$ \\ \hline
 $R_X(\frac{\pi}{2})_i \cdot CNOT_{i,j}$ & $Y_iZ_j,\  Y_iY_j,\  X_iX_j,\  Z_iX_j(-)$ \\ \hline
 $R_X(\frac{\pi}{2})_i \cdot R_Z(\frac{\pi}{2})_j \cdot CNOT_{i,j}$ & $Y_iX_j(-), Y_iZ_j,  X_iY_j(-), Z_iY_j$\\ \hline
\end{NiceTabular}
\egroup
\end{table}


\begin{figure}[b]
    \centering
    \includegraphics[width=0.4\textwidth]{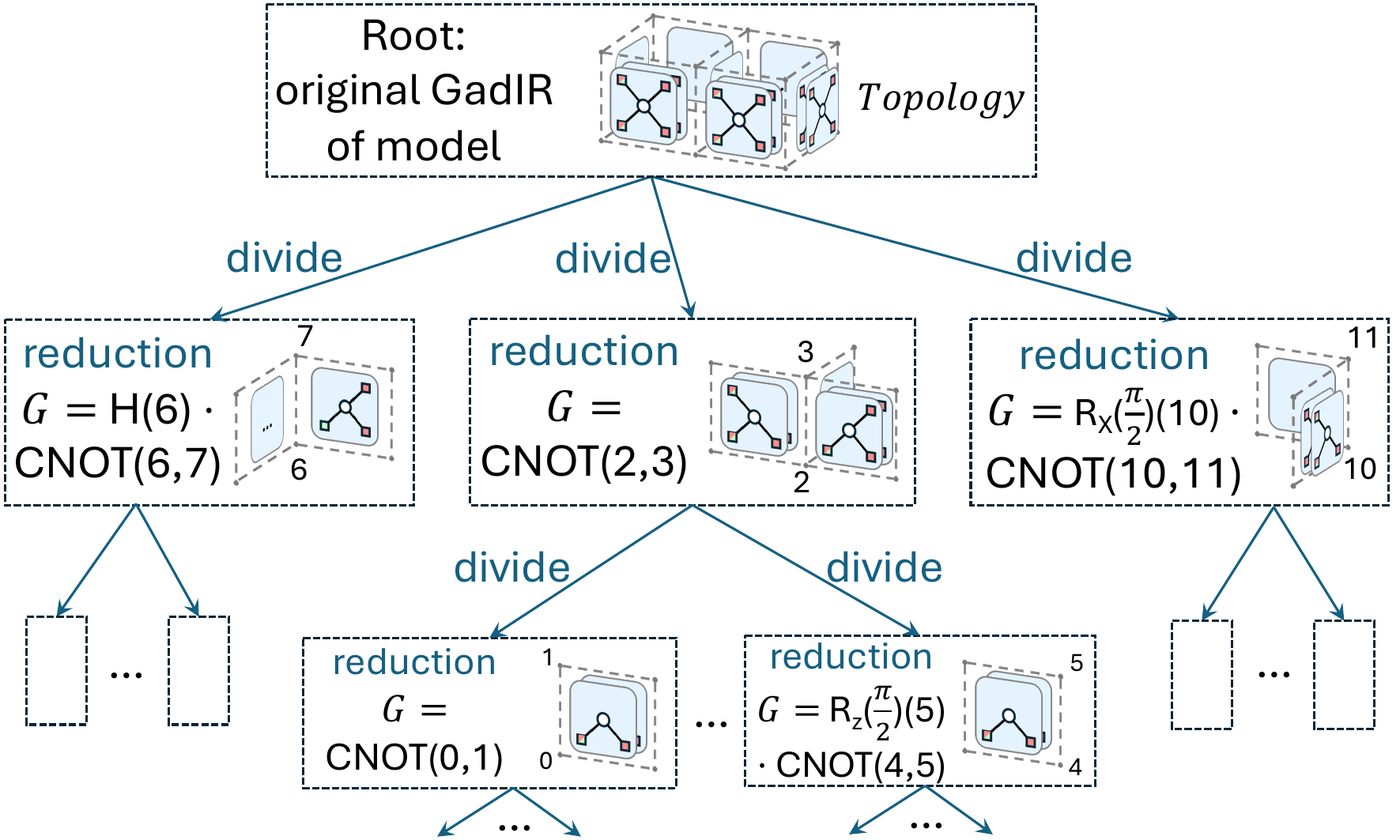}
    \caption{The gadget reduction tree.}
    \label{fig:reduction_tree}
\end{figure}
\hyphenation{topo-logy}
\subsection{Algorithm for Gadget Group Reduction} \label{algo_ggr}
We develop an algorithm to systematically apply reductions on the Hamiltonian of a general quantum many-body system, while trying to find the maximal reduction with the help of topology information.
Our algorithm is based on a tree structure, while its root stores GadIR of the original model.
The whole procedure corresponds to recursively growing the tree from its root layer by layer, and each reducible group of gadgets with a sub-topology are stored in the nodes, as illustrated in~\Cref{fig:reduction_tree}.
We define two operations for tree growth, which are \textit{groups dividing} and \textit{node updating}.

\textbf{Groups Dividing:} 
Given a set of Pauli gadgets with spatial topology, we divide them into multiple groups, while gadgets in the same group can be reduced by the same gate set $G$.
We use a greedy search to divide this set of gadgets into groups.
Specifically, we iterate over all the edges $\langle i,j \rangle$ in spatial topology $G_{topo}$, and for each $\langle i,j \rangle$ we iterate over all the $G$ in~\Cref{tab:reduction_rule}, counting the number of reducible gadgets for each $G$.
Throughout the iterations, we select the largest subset of gadgets which are reducible by a specific $G$ on specific topological edge, e.g. $G=\text{CNOT}{(2,3)}$.
This largest subset forms a new group, and we keep applying this operation to the other gadgets, until all are grouped.

\hyphenation{elimi-nating}
\textbf{Reduction and Updating:}
For a node, when all its gadgets are divided into groups, we spawn 
each of these groups as a child of this node.
Hence, each child stores: \circled{1} the GadIR of one divided group from its parent;
\circled{2} the specific gate set $G$ that has been selected to reduce all gadgets in this group.
Then we use this $G$ to reduce all gadgets, and update them into new reduced gadgets by eliminating their spiders (following rules in~\Cref{tab:reduction_rule}).

As illustrated in~\Cref{fig:reduction_tree}, starting from the root, we keep applying these two operations. When a node cannot find any reducible group of gadgets, it reaches the terminating condition and becomes a leaf node. 
The complete tree of GadIR is sent to compiler backend for further optimization.

\section{Compiler Backend Design}\label{sec:backend}
Compared to Paulihedral~\cite{li2022paulihedral}, 2QAN~\cite{lao20222qan} and Tetris~\cite{jin2024tetris} which tailor their Trotterization and scheduling algorithms for superconducting architecture, 
we expect that our compiler can perform optimization in these modules for arbitrary hardware architectures.
To achieve the goal, we design the Trotterization and scheduling algorithms directly on GadIR, which enable target-independent optimization.
We search commutation information between gadgets to schedule their simultaneous execution, thus enhancing parallelization and reducing circuit depth.
The compiler backend is implemented with PyZX~\cite{kissinger2019pyzx}, as the form of a ZX-diagram.



\subsection{Trotterization}
In the compiler frontend, GadIR serves as the representation of the Hamiltonian.
To execute the Hamiltonians on a quantum computer, we transform them into a ZX-diagram using the \textit{Suzuki-Trotter} method, as mentioned in~\Cref{subsec:qmbs}.
In this procedure which is called \textit{Trotterization}, 
each Hamiltonian term is individually extracted as ZX-diagram elements and appended to this circuit. 
Specifically for GadIR, each Hamiltonian term corresponds to a leaf node in the gadget reduction tree.

\begin{figure}[t]
    \centering
    \includegraphics[width=0.4\textwidth]{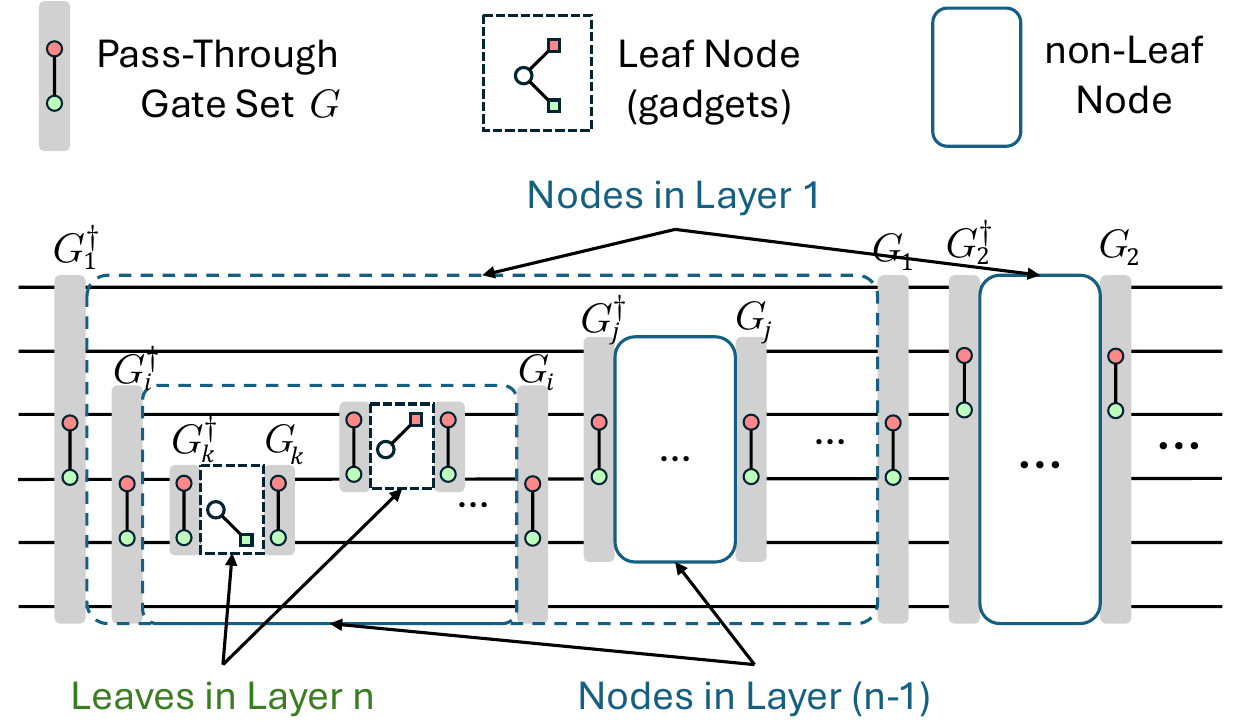}
    \caption{Trotterization: from gadget tree to ZX-diagram.}
    \label{fig:trotterization}
\end{figure}

Moreover, we have to add those extra gate sets $G$ used for gadget reductions, which are stored in each node of the tree.
Based on the gadget reduction procedure we introduced in~\Cref{subsec:ir_frontend}, applying reduction on a gadget group will insert adjoint gate sets $G^\dagger$ and $G$ on each side of the gadgets.
For a reduction tree that recursively applies group reduction to grow the nodes, correspondingly in the Trotterization circuit, the gate sets $G$ are recursively wrapped on both sides of the leaf-nodes (see~\Cref{fig:trotterization}).

\subsection{Scheduling}
\begin{figure}[b]
    \centering
    \includegraphics[width=0.4\textwidth]{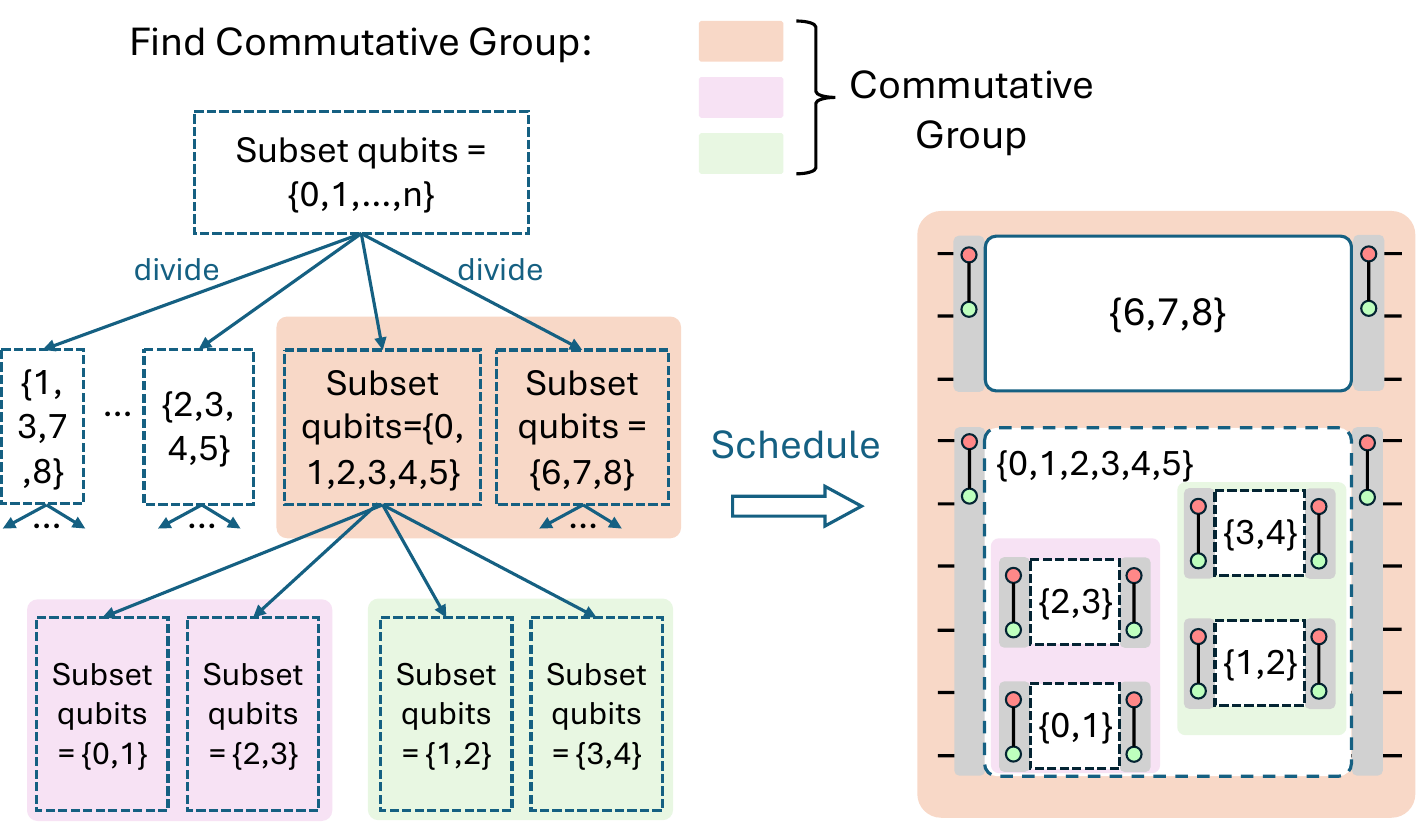}
    \caption{Scheduling: parallelizing the commutative groups.}
    \label{fig:scheduling}
\end{figure}

We arrange the time sequence of these leaf-nodes in Trotterization ZX-diagram via a scheduling method, as shown in~\Cref{fig:scheduling}.
This method relies on the topology subset information in each node of gadget reduction tree.
Looking into each node of~\Cref{fig:reduction_tree}, we can observe that each node and its descendants only involve a certain subset of the whole topology, and only perform operations on the qubits in this subset.
If the two subsets of two nodes have no intersection, they and their descendants are fully commutative with each other, so they can be scheduled to execute in parallel.

Hence we propose our scheduling strategy in~\Cref{fig:scheduling}:
starting from the root of reduction tree, in each node we store the subset information, which only contains the indices of qubits within this topology subset.
We accumulate the nodes from a layer (and from same parent) into groups, while in each group the subsets of nodes have no intersecting qubits.
Such accumulations are repeatedly applied until all nodes including leaves are merged into commutative group.
When we perform Trotterization, the nodes within same commutative group are arranged in the same layer of ZX-diagram.
\rev{Overall, our scheduling method exploits the parallelization of commutative groups in GadIR.}

\subsection{Hardware-Native Instruction Synthesis} \label{subsec:generate_ins}
Our final step is to synthesize native quantum programs specifically for each hardware architecture.
In this step, the spatial topology information that is preserved in GadIR is used for: 1. Assisting qubit initial placement for SC and Neutral-Atom; 2. Guiding ZX-diagram synthesis sequence for MBQC and FTQC.

\textit{Superconducting.}  
First, we extract the ZX-diagram into quantum circuit via the method described in \cite{duncan2020graph}.
Then, for qubit initial placement, we integrate a graph-based approach utilizing topology information from GadIR~\cite{bandic2023interaction,ren2024hardware}.
Specifically, we map the GadIR topology to a subgraph of the physical qubit coupling graph, while the graph edit distance (GED) between these two graphs is minimal.
The GED calculation uses the tool provided in~\cite{chang2020speeding,chang2022accelerating}.
After initial placement, the logical quantum circuit is synthesized to a physical circuit via the SABRE routing method~\cite{li2019tackling} in Qiskit~\cite{qiskit}.

\hyphenation{paralle-lize}
\hyphenation{app-lied}
\textit{Neutral-Atom.}
We use ZAC compiler~\cite{lin2025reuse} for synthesizing zoned-architecture neutral-atom quantum circuits, while customizing our topology-aware qubit initial placement.
For qubit initial placement, we use a min-cut algorithm to separate the topology in GadIR into
subgraphs.
The qubits from the same subgraph are mapped to the same SLM (spatial light modulator) row in storage zone.
Mapping them to the same SLM row can parallelize qubit movements~\cite{lin2025reuse} when applying CNOT gates among them.
This parallelization strategy leads to shorter execution time, thus reducing qubit decay.

\textit{FTQC with lattice surgery.}
First, we synthesize the sub-ZX-diagram in each leaf-node into an FTQC subroutine, using LaSsynth method~\cite{tan2024sat} implemented in Stim~\cite{gidney2021stim,lassynth}.
Then, we synthesize the pass-through Clifford gates (CNOT, H, etc.) via \Call{Block\_Synthesis}{} function in TQEC~\cite{tqec} (a QEC design package).
Next, we stitch all these subroutines using TQEC, and synthesize them into fault-tolerant operations such as lattice surgery.
The idea of synthesizing each subroutine separately avoids the massive overhead of synthesis.

\textit{Photonic MBQC.}
Emitter-based photonic MBQC is a novel scheme addressing the indeterminacy problem in MBQC~\cite{li2022photonic}.
We adopt the divide-and-conquer synthesis method from~\cite{ren2025scalable}, by synthesizing each leaf-node separately.
Specifically, the ZX-diagram in each leaf-node corresponds to a graph state for MBQC, and we synthesize the graph state generation scheme using GraphiQ~\cite{lin2024graphiq}.
Then these graph states are combined together using methods from \cite{ren2025scalable}.

\begin{figure*}[t]
\centering
    \centering
    {\includegraphics[width=0.85\textwidth]{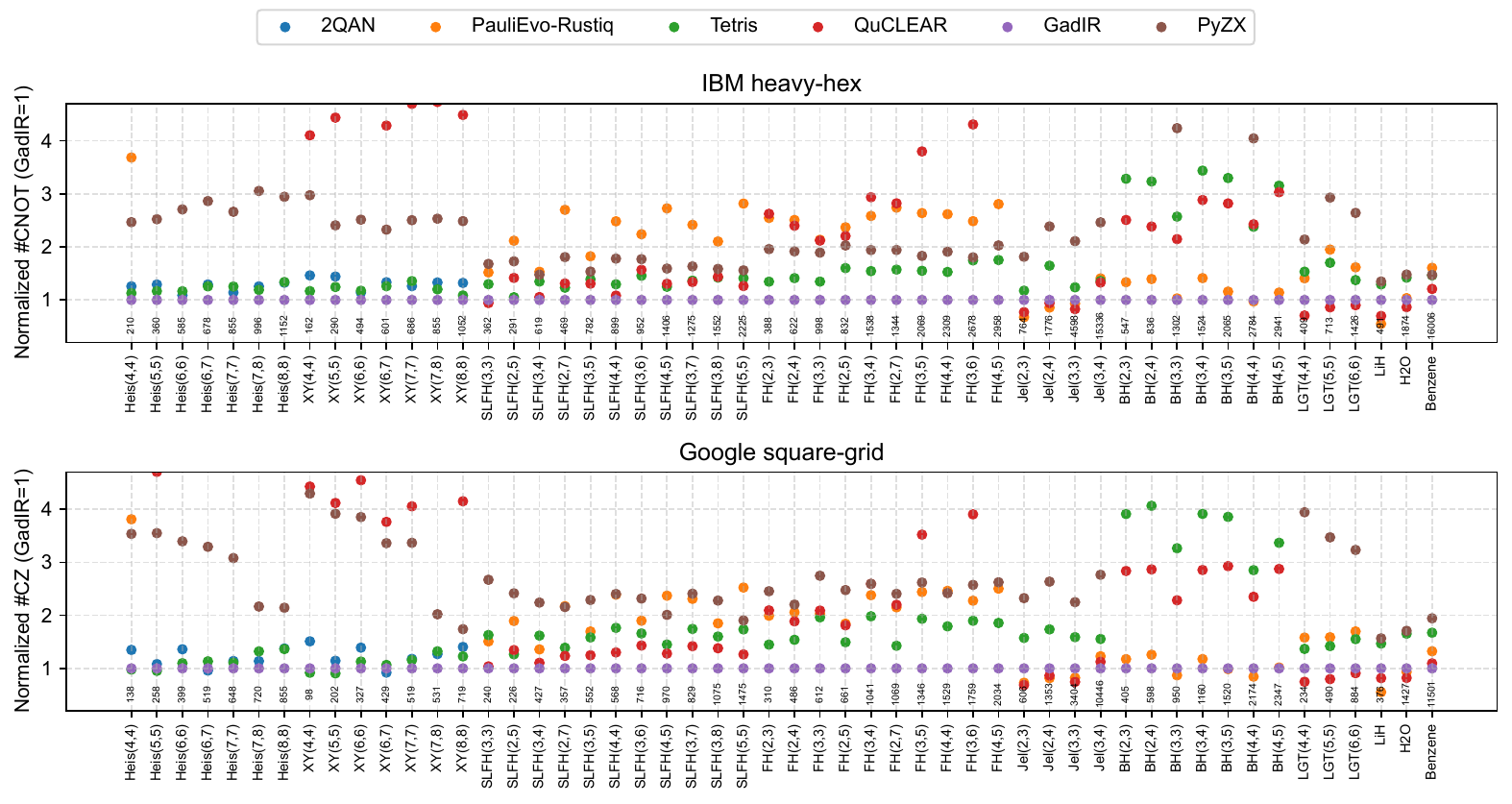}}
    \caption{\rev{Upper figure: The CNOT count for compilation to IBM quantum computer with heavy-hexagon qubits layout.
    Lower figure: The CZ count for compilation to Google quantum computer with square-grid qubits layout.}}
    \label{fig:res_sccnot}
\end{figure*}

\section{Experimental Methodology}\label{sec:methodology}
\hyphenation{experi-ment}
\subsection{Benchmarks}
We evaluate 8 different quantum many-body models in our experiments, each with a variety of model sizes.
These models cover all canonical model categories we mentioned: spin, fermionic, bosonic, and lattice gauge.
Apart from these, we add Molecule Hamiltonian models (LiH, H2O, Benzene) selected from ~\cite{liu2025quclear}, to demonstrate that our compiler is also compatible with quantum chemistry and variational quantum eigensolver (VQE) circuit. 
The models are listed in~\Cref{tab:benchmark}, with varying sizes defined by parameters $(x,y)$.
In the experiments, all the models are compiled with the second-order Trotterization with Suzuki-Trotter formula.
\begin{table}[h]
    \centering
    \caption{Benchmark information. In result figures they are tagged as: Heis=Heisenberg, SLFH=spinless Fermi-Hubbard, FH=spinful Fermi-Hubbard, Jel=Jellium, BH=Bose-Hubbard, LGT=Lattice Gauge Theory.}
    \label{tab:benchmark}
    \bgroup
    \def\arraystretch{1.1}
    \begin{tabular}{|c|c|c|c|}
        \hline
        Many-Body & Physical Model & Number \\
        Category &  Name & of Qubits \\ \hline \hline
        spin-qubit & Heisenberg$(x,y)$ & $x \times y$ \\ \hline
        spin-qubit & XY$(x,y)$ & $x \times y$ \\ \hline
        Fermion & spinless Fermi-Hubbard$(x,y)$ & $x \times y$ \\ \hline
        Fermion & spinful Fermi-Hubbard$(x,y)$ & $x \times y \times 2$ \\ \hline
        Fermion & Jellium$(x,y)$ & $x \times y$ \\ \hline
        Boson & Bose-Hubbard$(x,y)$ & $x \times y \times 2$ \\ \hline
        Lattice & (2+1)D Lattice   & $x \times y \times 2$ \\ 
        Gauge & $\mathbb{Z}_2$ Gauge Theory$(x,y)$ & $-x-y$\\ \hline
        Chemistry & LiH / H2O / Benzene & 6 / 8 / 12\\ \hline
    \end{tabular}
    \egroup
\end{table}


\subsection{Baselines and Metrics}
\rev{\textbf{Baselines:}} 
We choose several compilers from previous work as our baseline, which are \circled{1} QuCLEAR~\cite{liu2025quclear}, \circled{2} Tetris~\cite{jin2024tetris}, \circled{3} 2QAN~\cite{lao20222qan} and \circled{4} PauliEvo-Rustiq~\cite{paulievorustiq} (a Qiskit implementation integrating Paulihedral~\cite{li2022paulihedral} and Rustiq\cite{rustiq}). 
We note that 2QAN only supports spin-qubit models, because its IR and permutation-aware routing are designed for 2-local qubit Hamiltonians (e.g., ZZ and XX), while other models contain high-weight Hamiltonians (e.g., ZZZZ).
\circled{5} We implement an extra baseline (PyZX-opt) that directly uses PyZX on Pauli-string IR, and it also enables the circuit optimization passes in PyZX.
For MBQC and FTQC evaluation, the PyZX optimization is already integrated into baselines (QuCLEAR and Tetris), so we do not need to add the individual PyZX baseline.

\noindent\rev{\textbf{Metrics:}}
\circled{1} 
For SC architecture, the metrics are 2Q gate and circuit depth, aligning with the metrics used in baseline works~\cite{li2022paulihedral,lao20222qan,jin2024tetris,liu2025quclear}.
\circled{2} 
For NA architecture, apart from 2Q gate and circuit duration, we add the metric of fidelity. 
The main overhead of NA is moving the position of atoms to perform target gates, and the noise during movement is reflected in fidelity.
This aligns with the metric selection in SOTA compilers~\cite{wang2024atomique,lin2025reuse,huang2026zap}.
\circled{3} 
For emitter-based MBQC, the prevalent overhead is CZ operation between emitters, since it has significantly longer execution time and higher noise than other operations~\cite{lin2024graphiq}.
So we select the emitter CZ metric, aligning with SOTA~\cite{ren2025scalable,10.1145/3695053.3731085,lin2024graphiq}.
\circled{4} 
For FTQC architecture, the non-Clifford gates are approximately 100x slower than other operations due to state distillation~\cite{hao2025reducing,bravyi2005universal}. 
We select \#non-Clifford-gate as the metric, since it dominates the circuit execution overhead.
Another metric is the infidelity of FTQC circuit, simulated under a comprehensive circuit-level noise model.
These metrics align with SOTA FTQC compilers~\cite{hao2025reducing,hao2025compilation}.

\begin{figure}[tbp]
    \centering
    {\includegraphics[width=0.47\textwidth]{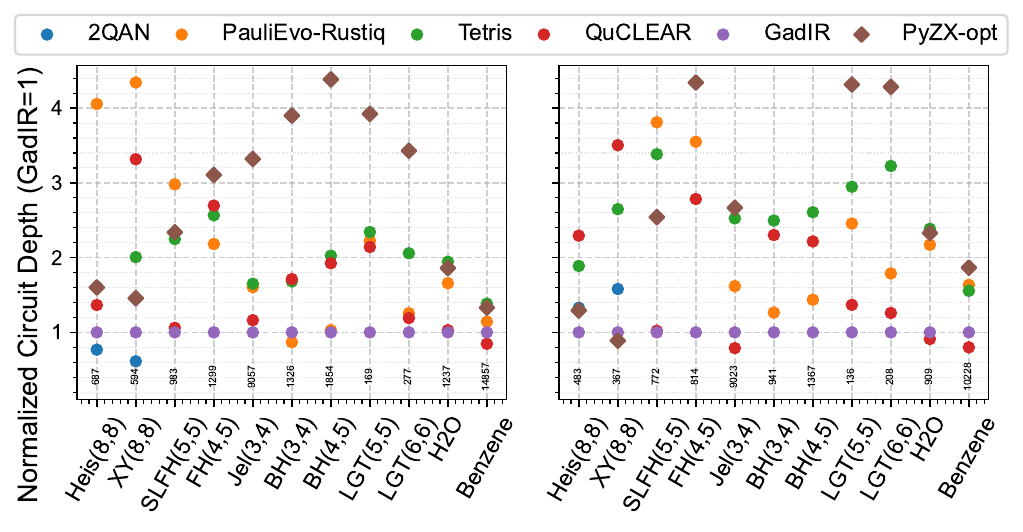}}
    \caption{\rev{Left figure: The circuit depth for compilation to IBM heavy-hexagon qubits layout. Right figure: The circuit depth for compilation to Google square-grid qubits layout.}}
    \label{fig:res_scdepth}
\end{figure}

\hyphenation{fidelity}
\begin{figure*}
    \centering
    {\includegraphics[width=0.82\textwidth]{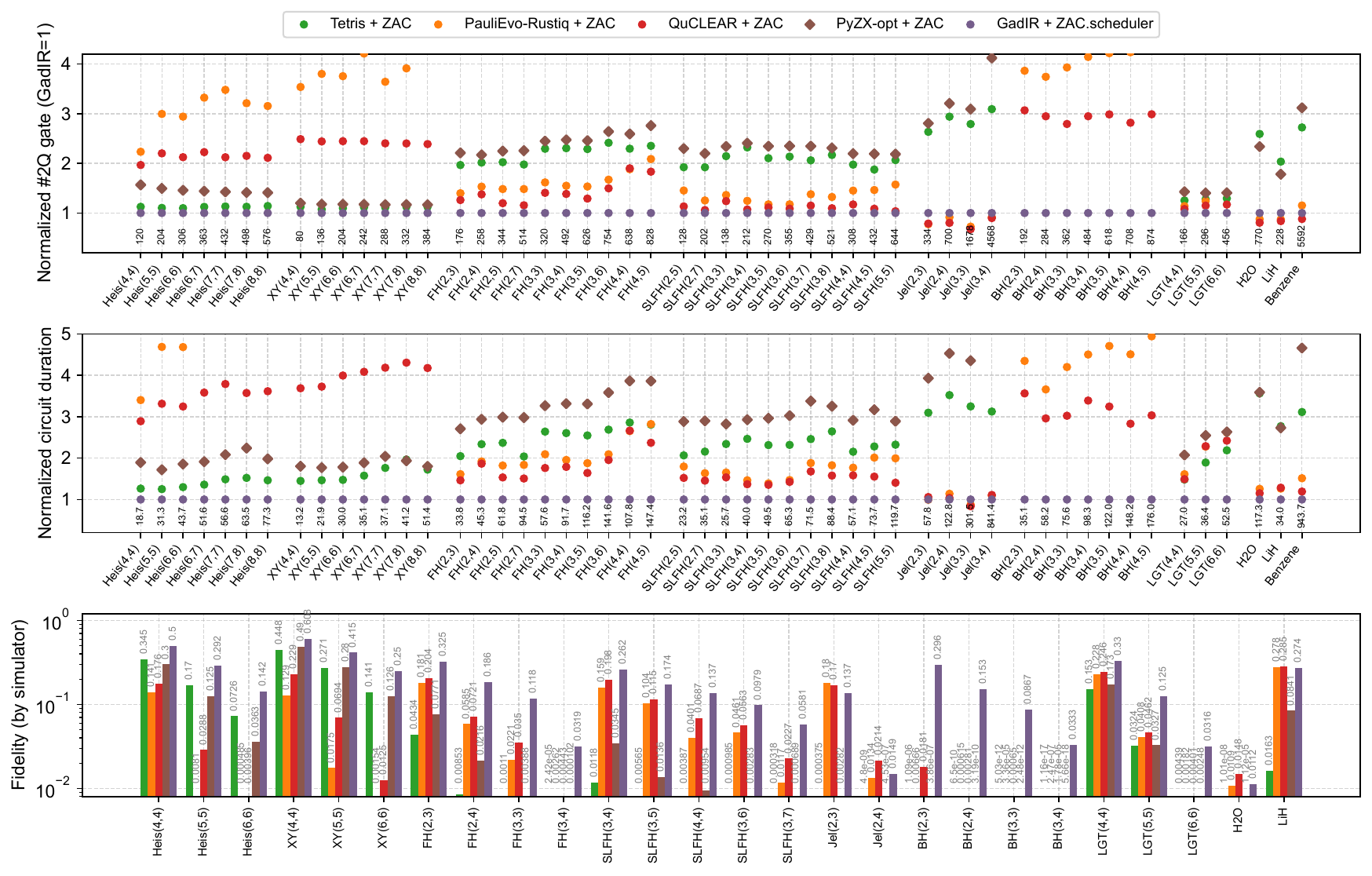}}
    \caption{Comparison of \#2Q gate (NA-native CZ), circuit duration, and fidelity on Neutral-Atom hardware.}
    \label{fig:res_na}
\end{figure*}

\section{Evaluation}\label{sec:eval}
\subsection{Superconducting Architecture}
We compare GadIR with all baseline compilers by counting the number of two-qubit gates (CZ) and circuit depth in compilation output.
Results for these two metrics are respectively shown in~\Cref{fig:res_sccnot} and~\Cref{fig:res_scdepth}.
On \textbf{IBM hardware} heavy-hexagon qubits layout, we reduce \textbf{\#CNOT} to 0.79$\times$ of 2QAN, 0.51$\times$ of PauliEvo-Rustiq, 0.69$\times$ of Tetris, 0.55$\times$ of QuCLEAR, and \rev{\textbf{0.46$\times$ of PyZX-opt}} on average;
and up to 0.68$\times$, 0.08$\times$, 0.29$\times$, 0.12$\times$, and \rev{\textbf{0.17$\times$}} respectively.
On \textbf{Google hardware} square-grid qubits layout, we reduce \textbf{\#CZ} to 0.83$\times$ of 2QAN, 0.56$\times$ of PauliEvo-Rustiq, 0.65$\times$ of Tetris, 0.57$\times$ of QuCLEAR, and \rev{\textbf{0.36$\times$ of PyZX-opt}} on average;
and up to 0.66$\times$, 0.09$\times$, 0.25$\times$, 0.14$\times$, and \rev{\textbf{0.13$\times$}} respectively.
Due to the page limit, we only select part of the results for circuit depth comparison in~\Cref{fig:res_scdepth}.
It can be observed that GadIR outperforms all baselines in most of the benchmarks, and within each benchmark the reduction rate improves with increasing model size.
The trend of improvement implies that spatial topology information of GadIR becomes more important when many-body models scale up.

\subsection{Neutral-Atom Architecture}
We compare GadIR with baselines on zoned-architecture \textbf{neutral atom} platform.
The baseline compilers are integrated with ZAC~\cite{lin2025reuse} with the following pipeline:
We run the compilation with an all-to-all qubit layout and get the output as a logical circuit, then pass the circuit to ZAC for atom initial placement and scheduling.
Meanwhile, our compiler defines customized initial placements and only uses the atom scheduling module of ZAC, as introduced in~\Cref{subsec:generate_ins}.
\rev{
We use the following parameters: Rydberg 2Q gate time is 0.36 us, 1Q gate time is 52 us, atom-transfer activation/deactivation time is 15 us, 2Q/1Q/atom-transfer fidelities are 0.996, 0.9998, and 0.9995, respectively, and the qubit coherence time is 4e6 us.
Circuit fidelity is then modeled as the product of 1Q gate fidelity, Rydberg 2Q gate fidelity, atom-transfer fidelity, and coherence survival over the scheduled circuit duration.
}

The results of 2Q gate, duration and fidelity are shown in~\Cref{fig:res_na}. 
GadIR reduces \textbf{\#2Q gate} to 0.54$\times$ of Tetris, 0.59$\times$ of PauliEvo-Rustiq, 0.71$\times$ of QuCLEAR, and 0.48$\times$ of PyZX-opt on average; for \textbf{circuit duration}, GadIR reduces duration to 0.44$\times$, 0.45$\times$, 0.52$\times$, and 0.34$\times$ of those same baselines on average.
GadIR achieves \textbf{fidelity} improvements of 13.7$\times$ over Tetris, 11.2$\times$ over PauliEvo-Rustiq, 6.8$\times$ over QuCLEAR, and 10.9$\times$ over PyZX-opt on average.
In the results, every baseline compiler demonstrates unexpectedly poor performance on certain benchmark models, highlighting the importance of robustness across diverse hardware.

\subsection{Early-Fault-Tolerant Architecture}
\begin{figure}[tbp]
    \centering
    \includegraphics[width=0.4\textwidth]{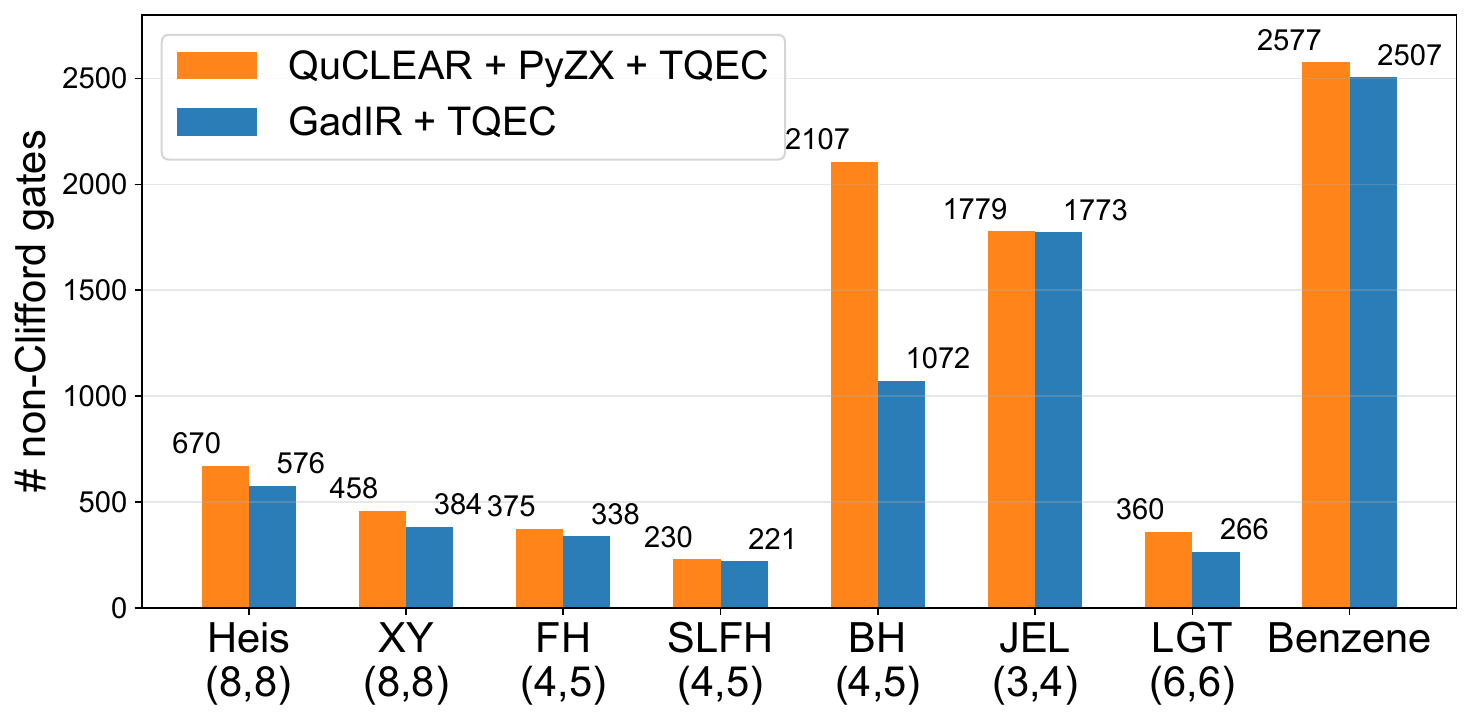}
    \caption{Numbers of non-Clifford gates on FTQC.}
    \label{fig:noncliff}
\end{figure}

We compare GadIR with baselines on \textbf{FTQC} platform with surface code.
Limited by massive computation overhead for FTQC simulator, only the latest SOTA compiler -- QuCLEAR~\cite{liu2025quclear} is selected as baseline in our experiments.
We integrate QuCLEAR with PyZX~\cite{kissinger2019pyzx} to convert compilation output into ZX-diagrams, and synthesize the lattice surgery scheme via TQEC~\cite{tqec}.
As shown in \Cref{fig:noncliff}, GadIR reduces the number of non-Clifford gates
compared to QuCLEAR.

\begin{figure}[tbp]
    \centering
    {\includegraphics[width=0.45\textwidth]{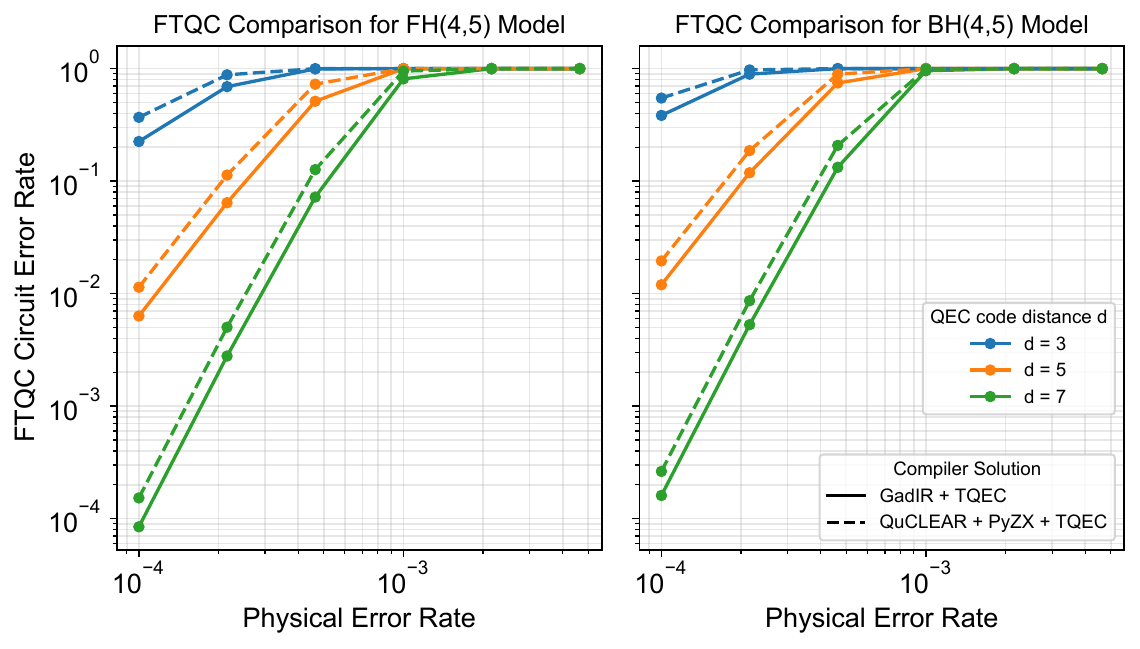}}
    \caption{\rev{Comparison of circuit infidelity on FTQC.}}
    \label{fig:res_ftqc}
\end{figure}

We compare the \textbf{infidelity} of the FTQC circuits by running them on Stim simulator with PyMatching QEC decoder~\cite{Higgott2025sparseblossom}.
For each physical error rate $p$, we use the circuit-level SI1000 noise model~\cite{gidney2021fault} with CZ error rate $p$, 1Q Clifford gate $p/10$, measurement $5p$, reset $2p$, idle $p/10$, and resonator-idle $2p$.
Aligning with SOTA early-FTQC compilers~\cite{tan2024sat,hao2025compilation}, the qubit plane is a dynamic 2D nearest-neighbor lattice-surgery patch layout~\cite{tan2024sat,low2026denser} generated by \textsc{LaSsynth}~\cite{lassynth}, rather than a fixed data-ancilla layout.
We model the non-Clifford rotations by applying execution noise to the resulting T gates using a depolarizing error~\cite{bravyi2005universal,hao2025compilation}.
The results (code distance = 3, 5, 7) are given in~\Cref{fig:res_ftqc}.
GadIR reduces the \textbf{logical CNOT} subroutines to \textbf{0.58× of QuCLEAR} on average, which leads to a \textbf{0.64× average circuit infidelity} in the non-saturated regime and \textbf{0.58×} in the low-error regime.

\vspace{-2mm}
\vspace{-2mm}

\subsection{Emitter-Based MBQC Architecture}
\begin{figure}[tbp]
    \centering
    {\includegraphics[width=.46\textwidth]{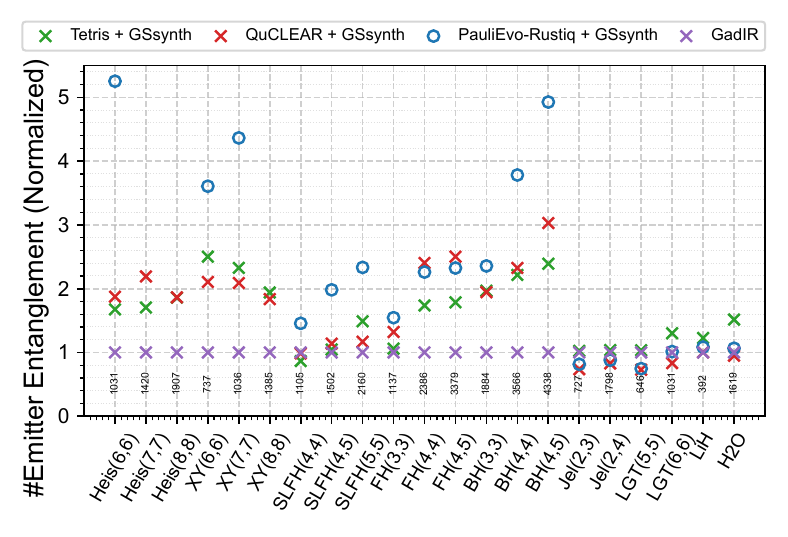}}
    \caption{\rev{The number of emitters CZ operations for compilation to emitter-based MBQC hardware.}}
    \label{fig:res_emitter}
\end{figure}

We compare GadIR with the baselines on emitter-based MBQC platform, while using silicon-dot photon emitter~\cite{larocque2024tunable,zhu2025hybrid,cogan2023deterministic} as synthesis target.
The baseline compilers are integrated with SOTA graph-state synthesis algorithm (GSsynth) from \cite{takou2025optimization}.
In the experiments, we compare the number of \textbf{emitter entanglement operations} (emitter-emitter CNOT), which is much more expensive than other operations~\cite{li2022photonic}.
The results in~\Cref{fig:res_emitter} show that GadIR reduces the entanglement overhead to \rev{0.69$\times$} that of Tetris+GSsynth, \rev{0.75$\times$} that of QuCLEAR+GSsynth, and \rev{0.56$\times$} that of PauliEvo-Rustiq+GSsynth on average.

\subsection{Runtime and Scalability}
\begin{figure}[t]
    \centering
    \includegraphics[width=0.38\textwidth]{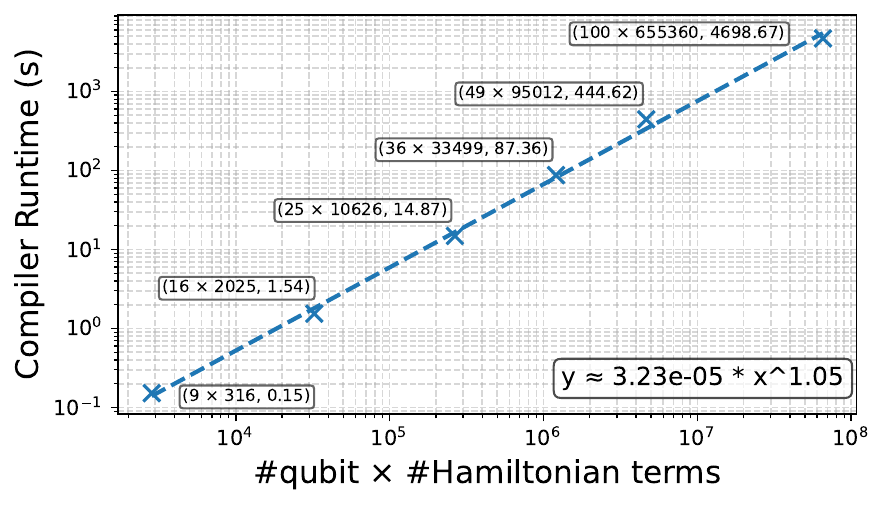}
    \caption{GadIR runtime scaling \rev{with model size}.}
    \label{fig:runtime}
\end{figure}

Here we analyze the scalability of our algorithms by measuring their runtime.
For Jellium models, we record the total execution time of the compiler pipeline, including gadget reduction, Trotterization, scheduling, while synthesis time is excluded since it depends on selection of known SOTA methods.
The results are given in~\Cref{fig:runtime}, where it can be observed that the runtime scales nearly linearly with the parameter $x=$ \#qubit $\times$ \#Hamiltonian, giving the scaling of compilation runtime $t\approx3e-5 \times x^{1.05}$. 
\rev{Moreover, we perform a smoke test for large-scale model on FH(30, 30), yielding a $9.1\times10^3$ s runtime, matching the scaling curve.}
This result implies the scalability of our compiler framework, and its practicality for median- to large-scale quantum algorithm.

\begin{figure}[t]
    \centering
    {\includegraphics[width=0.45\textwidth]{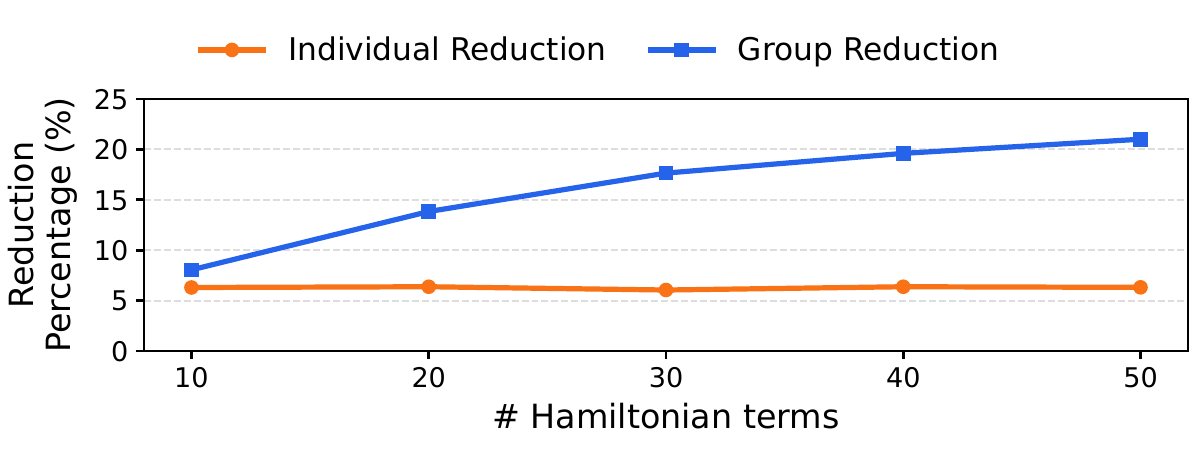}}
    \caption{\rev{Comparison of individual gadget reduction~\cite{cowtan2019phase} with our gadget group reduction, on random Hamiltonian.}}
    \label{fig:isolated}
\end{figure}

\rev{\subsection{Individual vs. Group Reduction}}
\rev{
The prior work~\cite{cowtan2019phase,cowtan2020generic} extracts Clifford gates from each gadget and passes it through all other gadgets to eliminate spiders, but it often creates extra spiders thus undermining the effectiveness of gadget reduction.
}
\rev{Here we perform an experiment to emphasize the broad impact of gadget group reduction, comparing to the individual gadget reduction introduced in \cite{cowtan2019phase,cowtan2020generic}.
In \Cref{fig:isolated}, we compare individual and group reduction on different sizes of Hamiltonian terms. 
Each size is evaluated with 1000 randomly generated instances of Hamiltonian, and we calculate the average reduction percentages.
Note that in this experiment, we modify the group reduction method to ignore  topology information and only use a topology-agnostic search strategy.
The group reduction method outperforms individual reduction when the size of Hamiltonian scales up.
Hence, the non-trivial improvement of group reduction itself provides better scalability for large-scale Hamiltonian models.
}

\begin{table*}[t]
\centering
\caption{\rev{
Ablation studies. The table reports relative change in 2Q gate count and circuit depth (duration) when each module is disabled. \ul{Since NA architecture provides an all-to-all physical qubit connectivity, the 2Q gate count remains unchanged. For NA, the backend modules mainly reduce atom movement distance, thereby reducing circuit duration.}}}
\label{tab:ablation_study}
\begingroup
\color{black}
\arrayrulecolor{black}
\fontsize{8pt}{10pt}\selectfont
\begin{tabular}{|l|c|c|c|c|c|c|c|c|}
\hline
\multirow{2}{*}{Architecture}
& \multicolumn{2}{c|}{post-Trotterization gate cancelling}
& \multicolumn{2}{c|}{w/o HW-native synthesis}
& \multicolumn{2}{c|}{w/o entire GadIR backend}
& \multicolumn{2}{c|}{w/o topology-preserving} \\
\cline{2-9}
& \quad \# 2Q gate \:\quad  & Depth
& \# 2Q gate & Depth
& \# 2Q gate & Depth
& \# 2Q gate & Depth \\
\hline
SC (Google)  & -8.5\% & +19.1\% & +15.6\% & +8.7\%  & +17.8\% & +46.5\% & +43.6\% & +61.2\% \\
\hline
SC (IBM)     & -7.3\% & +15.5\% & +14.4\% & +6.3\%  & +17.0\% & +53.3\% & +37.5\% & +38.2\% \\
\hline
Neutral-Atom & -4.8\% & +24.0\% & 0       & +22.4\% & 0       & +68.3\% & +28.3\% & +69.0\% \\
\hline
\end{tabular}%
\endgroup
\end{table*}

\begin{figure}[t]
    \centering
    \includegraphics[width=0.45\textwidth]{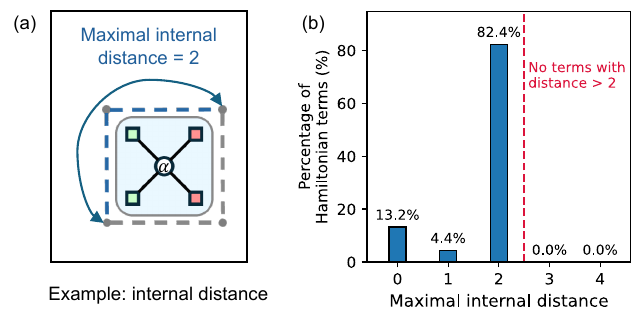}
    {\includegraphics[width=0.45\textwidth]{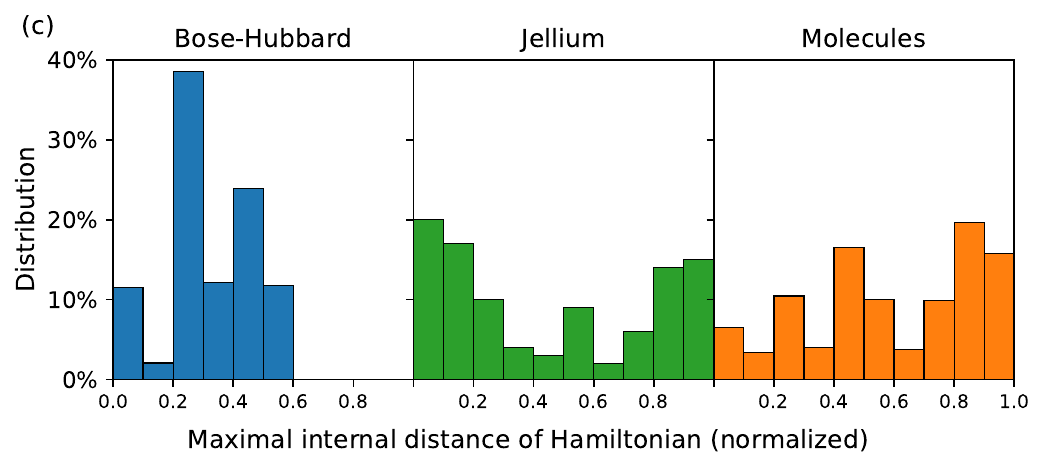}}
    {\includegraphics[width=0.45\textwidth]{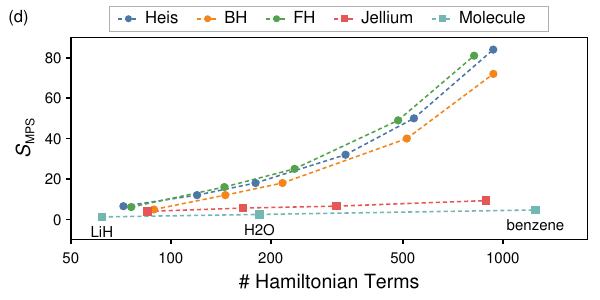}}
    \caption{
    (a) Illustration of how the maximal internal distance is counted for a representative Hamiltonian term. 
    (b) Distribution of maximal internal distance for all Hamiltonians in a $2\times3$ bosonic model.
    \rev{(c) The distribution of \textit{maximal internal distance} of the Hamiltonian terms from different models.}
    \rev{(d) The $S_{MPS}$ metric for different types of many-body system, which dominates the classical simulation overhead of their Hamiltonian models.}}
    \label{fig:locality_analysis}
\end{figure}

\subsection{Hamiltonian Locality Analysis} \label{sec:locality_analysis}
Here we explore the locality of Hamiltonian in quantum many-body systems, serving as the motivation for the spatial topology-preserving design of GadIR.
We introduce the \textit{maximal internal distance} as a metric for quantifying the locality of a Hamiltonian term. 
This metric is defined as the maximum graph distance between any pair of qubits in the Hamiltonian term, where the distance is measured as the length of the shortest edge path between them on the GadIR topology graph.
An illustrative example of maximal internal distance is given in \Cref{fig:locality_analysis}a.
\ul{When the \textit{maximal internal distances} of Hamiltonian terms in a model are generally smaller, the model has stronger topological constraints, implying stronger locality of Hamiltonian in this model.}

We calculate the distribution of \textit{maximal internal distance} for a $2\times3$ bosonic model, and results are shown in \Cref{fig:locality_analysis}b.
In this model, all the Hamiltonian terms have a \textit{maximal internal distance} $\leq2$.
Observing the distribution, we know that qubit interactions in this model are constrained to short distances.
Based on these constraints, we emphasize the locality of quantum many-body systems.
The locality benefits our gadget group reduction method, and details are introduced in \Cref{sec:frontend}.
The utilization of locality justifies preserving spatial topology in GadIR.

\rev{
From the evaluation results on SC and NA architectures, we notice that GadIR has more significant improvement on quantum many-body system benchmarks, than on Jellium and Molecular benchmarks.
The differences are explained by the stronger Hamiltonian locality in these quantum many-body systems.
In \Cref{fig:locality_analysis}, we analyze the average distribution of \textit{maximal internal distance} of Hamiltonian terms.
The distribution of BH models is concentrated around lower distances (implying strong topology-constraints), while the distribution of Molecule models tends to be more uniform (implying weak topology-constraints). 
}

\rev{
We observe an empirical correlation between Hamiltonian locality and the classical simulation overhead of many-body systems, as shown in \Cref{fig:locality_analysis}.
We quantify the overhead using MPS entropy $S_{\mathrm{MPS}}$~\cite{schollwoeck2011density,cirac2021matrix}. 
The classical simulation complexity~\cite{vidal2003efficient,schuch2008entropy} grows exponentially with $S_{\mathrm{MPS}}$:
$$T_{classical} \sim poly(N_{qubit})~e^{S_{\mathrm{MPS}}}.$$
We calculate $S_{\mathrm{MPS}}$ using MPS entanglement profiling~\cite{schollwoeck2011density,orus2014practical}.
The models where GadIR achieves larger reductions exhibit faster growth of $S_{\mathrm{MPS}}$ as the system size increases. This suggests that, the strong-locality models targeted by GadIR are associated with higher classical simulation overhead than the molecular models~\cite{leclerc2026one}.
}

\rev{\subsection{Ablation Studies}}
\rev{
We perform four ablation studies with results in \Cref{tab:ablation_study}. 
(1) We modify the backend for maximizing gate cancellation after Trotterization, by replacing the scheduling strategy for $G$ with the method in Paulihedral~\cite{li2022paulihedral}. The new backend slightly reduces 2Q gates but increases circuit depth. While the cancellation chances for CNOTs are $G$ is limited, our GadIR backend focuses on capturing parallelization for commutative groups.
We replace the (2) hardware-native synthesis or (3) the entire backend with Qiskit compiler. For SC, HW-native synthesis contribute more on reducing 2Q gate while the other modules contribute to reducing depth. For NA, each module has reasonable reduction on circuit duration.
(4) By ablating the topology, we can observe that it is important for GadIR optimization, by capturing significantly more gadget reduction chances.
}

\section{Related Work}

\ul{Kernpiler}~\cite{decker2025kernpiler} groups commutative Hamiltonian terms, and use Monte Carlo searching to find a circuit approximating each Hamiltonian group.
Kernpiler focuses on optimizing the Trotterization stage, while our compiler focuses on reducing Hamiltonian terms before Trotterization.
Its approximation method is orthogonal to our gadget reduction optimization and can be integrated to  reduce Trotterization error.
\ul{HATT}~\cite{liu2024fermihedral,liu2025hatt} proposes novel methods for fermionic-to-qubit encoding.
While our work simply uses Bravyi-Kitaev (BK) encoding for its fast runtime, these encoding algorithms can replace BK further to reduce the number of Pauli operators.

\ul{Rustiq}~\cite{rustiq} focuses on greedily synthesizing flattened Pauli-rotation sets into low-CNOT/depth circuits through Clifford basis changes. By contrast, our compiler preserves the spatial topology and discovers optimizations from Pauli gadget reduction.

\ul{PHOENIX}~\cite{yang2025phoenix} introduces a Pauli-level optimization that uses Clifford transformations to simultaneously reduce weights of Pauli strings in binary symplectic form representation. 
Unlike PHOENIX~\cite{yang2025phoenix}, which relies on sequential Clifford transformations to simplify groups of flattened Pauli strings, our frontend operates on topology-preserving Pauli gadgets, enabling reduction opportunities to be discovered directly along physical-topology edges. 

\ul{Genesis}~\cite{chen2023speed} is a compiler for Fermion-Boson Hamiltonian on continuous-variable discrete-variable (CV-DV) quantum architecture.
Our insight of preserving spatial topology of bosonic models can be integrated into CV-DV compiler, thus one of our future directions is extending GadIR to support its synthesis.
\ul{PhasePoly}~\cite{chen2025phasepoly} explores the phase-polynomial optimization in quantum circuit. 


Optimizing quantum simulation and variational quantum algorithms is a topic that has drawn significant research attention~\cite{kim2025distribution,wang2023efficient,wang2024qoncord,wu2021mitigating,wu2021towards,seifert2024clapton,bhattacharyya2023optimal,zhang2023disq,dangwal2023varsaw,ravi2023navigating,ravi2022cafqa,ravi2022vaqem,meher2024error, BQSim, Taskflow, Q-TranSim,ren2024coherent}, while we focus on quantum many-body systems simulation and use its topology information as a heuristic.
A series of previous works have been discussing the importance of architecture-aware compilation~\cite{10.1145/3445814.3446750,10.1145/3307650.3322213,10.1145/3695053.3731085,10.1145/3622781.3674179,10.1145/3622781.3674185,Jang2024recompiling,kim2025qr,roy2025forensics,kundu2025inverse,das2025optimization,hua2023caqr,murali2020architecting,murali2019noise,javadiabhari2014scaffcc,murali2019formal,tang2021cutqc,tang2022scaleqc,tang2025tensorqc,tornow2025qvm,giortamis2025qos,lao2021designing,murali2019full,zzxtalk,ren2026photonic,ren2024machine,zhang2023decoder,zhong2026mera,zhong2026qutuner,yang2026unifying,kong2026directional,huang2025p2p}, while we try to address this problem with the novel GadIR design.

\section{Conclusion}
We propose a modularized frontend-backend compiler for quantum many-body systems simulation. We introduce GadIR to preserve spatial-topology, enabling gadget group reduction in the frontend, and enabling backend synthesis for hardware-native programs across four different architectures. Evaluations on all canonical models show significant reductions in dominant overheads.

\clearpage
\bibliographystyle{ACM-Reference-Format}
\bibliography{references}

\clearpage
\newpage
\appendix
Appendix is not allowed in MICRO26 submission!!! Remove later
\section{ZX-calculus}
\label{sec:appendix_zx}
Here we give the derivation of CNOT in ZX-calculus:
\begin{equation}\label{cnotdecomp}
\arraycolsep=1pt\def\arraystretch{0.6}
\begin{array}{rcl}
    \scalebox{0.9}{\begin{tikzpicture}
	\begin{pgfonlayer}{nodelayer}
		\node [style=Z dot] (0) at (0, 0) {};
		\node [style=none] (1) at (-1, 0) {};
            \node [style=none] (2) at (1, 0.75) {};
            \node [style=none] (3) at (1, -0.75) {};
        \end{pgfonlayer}
        \begin{pgfonlayer}{edgelayer}
		\draw (1.center) to (0);
            \draw [bend left=30] (0) to (2);
            \draw [bend right=30] (0) to (3);
        \end{pgfonlayer}
\end{tikzpicture}} \quad & = & \quad \ketbra{00}{0} + \ketbra{11}{1} \\
    \\
    \scalebox{0.9}{\begin{tikzpicture}
	\begin{pgfonlayer}{nodelayer}
		\node [style=X dot] (0) at (0, 0) {};
		\node [style=none] (1) at (1, 0) {};
            \node [style=none] (3) at (-1, 0.75) {};
            \node [style=none] (2) at (-1, -0.75) {};
        \end{pgfonlayer}
        \begin{pgfonlayer}{edgelayer}
		\draw (1.center) to (0);
            \draw [bend left=30] (0) to (2);
            \draw [bend right=30] (0) to (3);
        \end{pgfonlayer}
\end{tikzpicture}} \quad & = & \quad \ketbra{+}{++}\  + \ketbra{-}{--}
\end{array}
\end{equation}
\vspace{2mm}
\begin{equation}\label{cnot}
    \begin{tikzpicture}
	\begin{pgfonlayer}{nodelayer}
		\node [style=Z dot] (0) at (0, 0.5) {};
		\node [style=none] (1) at (1.25, 0.5) {};
		\node [style=X dot] (2) at (0, -0.5) {};
		\node [style=none] (3) at (-1.25, 0.5) {};
		\node [style=none] (4) at (-1.25, -0.5) {};
		\node [style=none] (5) at (1.25, -0.5) {};
	\end{pgfonlayer}
	\begin{pgfonlayer}{edgelayer}
		\draw (0) to (2);
		\draw (0) to (3.center);
		\draw (0) to (1.center);
		\draw (5.center) to (2);
		\draw (2) to (4.center);
	\end{pgfonlayer}
\end{tikzpicture} \quad =\quad 
    CNOT := \ 
    \left(\begin{matrix}
      1 & 0 & 0 & 0 \\
      0 & 1 & 0 & 0 \\
      0 & 0 & 0 & 1 \\
      0 & 0 & 1 & 0 \\
    \end{matrix}\right)
\end{equation}

Here is an simple example of the relationship between ZX-diagram and quantum circuit.
\begin{figure}[h!]
\centering
\begin{subfigure}{.205\textwidth}
    \centering
    \includegraphics[width=1\textwidth]{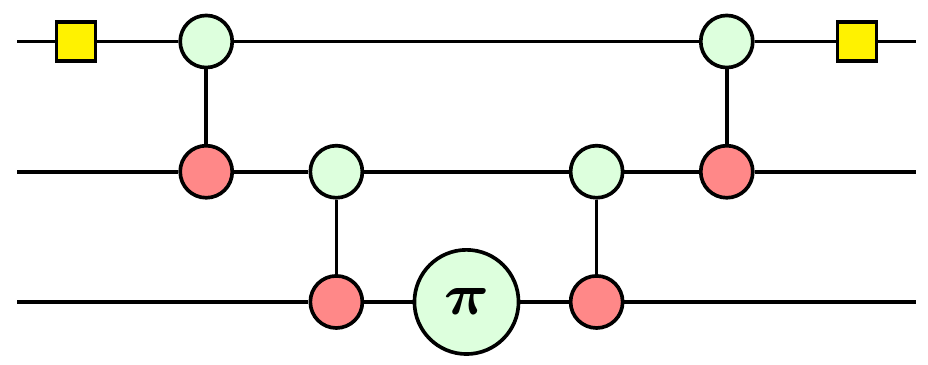}
\end{subfigure}%
\hfill
\begin{subfigure}{.25\textwidth}
    \centering
    \includegraphics[width=1\textwidth]{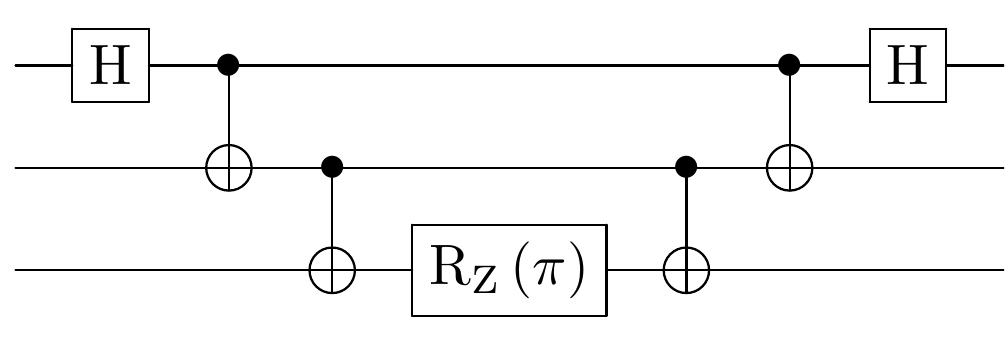}
\end{subfigure}
\caption{An example of ZX-diagram (left) and the equivalent quantum circuit (right).}
\label{fig:examples}
\vspace{-2mm}
\end{figure}

Here are the approach for transforming Pauli gadget into normal ZX-diagram elements.
\circled{1} decomposing Y-spiders to Z-spider sandwiched by single-qubit rotations $R_X$, and \circled{2} decomposing the coefficient box with a X-spider connected to a dangling Z-spider:
\begin{equation}
\scalebox{0.85}{
    \def\scl{0.7}
\begin{tikzpicture}[baseline={(2)}]
	\begin{pgfonlayer}{nodelayer}
        \node [style=none] (14) at (-0.5, 0*\scl) {$q_3$};
        \node [style=none] (15) at (-0.5, 1.5*\scl) {$q_2$};
        \node [style=none] (16) at (-0.5, 3*\scl) {$q_1$};
        \node [style=none] (17) at (-0.5, 4.5*\scl) {$q_0$};
		\node [style=none] (0) at (0, 0 *\scl) {};
		\node [style=none] (1) at (0, 1.5 *\scl) {};
		\node [style=none] (2) at (0, 3 *\scl) {};
		\node [style=none] (3) at (0, 4.5 *\scl) {};
		\node [style=Zexp] (4) at (1.5, 0 *\scl) {};
		\node [style=Xexp] (5) at (1.5, 4.5 *\scl) {};
		\node [style=Yexp] (6) at (1.5, 3 *\scl) {};
		\node [style=none] (9) at (5, 4.5 *\scl) {};
		\node [style=none] (10) at (5, 3 *\scl) {};
		\node [style=none] (11) at (5, 1.5 *\scl) {};
		\node [style=none] (12) at (5, 0 *\scl) {};
		\node [style=Aexp] (13) at (3.5, 2 *\scl) {$\alpha$};
	\end{pgfonlayer}
	\begin{pgfonlayer}{edgelayer}
		\draw (3.center) to (9.center);
		\draw (2.center) to (10.center);
		\draw (1.center) to (11.center);
		\draw (12.center) to (4);
		\draw (4) to (0.center);
		\draw (5) to (13);
		\draw (13) to (6);
		\draw (4) to (13);
	\end{pgfonlayer}
\end{tikzpicture} \qquad = \qquad \def\scl{0.7}
\begin{tikzpicture}[baseline={(2)}]
	\begin{pgfonlayer}{nodelayer}
		\node [style=none] (0) at (-0.5, 0 *\scl) {};
		\node [style=none] (1) at (-0.5, 1.5 *\scl) {};
		\node [style=none] (2) at (-0.5, 3 *\scl) {};
		\node [style=none] (3) at (-0.5, 4.5 *\scl) {};
		\node [style=Z dot] (4) at (1.5, 0 *\scl) {};
		\node [style=X dot] (5) at (1.5, 4.5 *\scl) {};
		\node [style=Z dot] (6) at (1.5, 3 *\scl) {};
		\node [style=none] (9) at (5.5, 4.5 *\scl) {};
		\node [style=none] (10) at (5.5, 3 *\scl) {};
		\node [style=none] (11) at (5.5, 1.5 *\scl) {};
		\node [style=none] (12) at (5.5, 0 *\scl) {};
		\node [style=X dot] (13) at (3.5, 0.75 *\scl) {};
            \node [style=X phase dot] (14) at (0.5, 3 *\scl) {$\frac\pi2$};
            \node [style=X phase dot] (15) at (3.5, 3 *\scl) {$-\frac\pi2$};
            \node [style=Z phase dot] (16) at (4.75, 0.75 *\scl) {$2\alpha$};
	\end{pgfonlayer}
	\begin{pgfonlayer}{edgelayer}
		\draw (3.center) to (9.center);
		\draw (2.center) to (10.center);
		\draw (1.center) to (11.center);
		\draw (12.center) to (4);
		\draw (4) to (0.center);
		\draw (5) to (13);
		\draw (13) to (6);
		\draw (4) to (13);
            \draw (13) to (16);
	\end{pgfonlayer}
\end{tikzpicture}}
\end{equation}

\vspace{3mm}
There are various rewriting rules for ZX-diagram, while the basic rule is: spider with same color are commutative, thus they can pass through each other on the wire. 
In addition, two neighboring spiders with same color can be merged into a new one, with new the coefficient being the sum of each coefficient of them:
\begin{equation}\label{basic_rule}
    \scalebox{0.85}{\begin{tikzpicture}
	\begin{pgfonlayer}{nodelayer}
		\node [style=Z phase dot] (0) at (-1.5, -0.625) {$\beta$};
		\node [style=none] (1) at (-0.5, -0.375) {};
		\node [style=none] (2) at (-3.25, -0.375) {};
		\node [style=none] (3) at (-3.25, -1.625) {};
		\node [style=none] (4) at (-0.5, -1.625) {};
		\node [style=none, rotate=90] (7) at (-3.5, -1) {...};
		\node [style=none, rotate=90] (8) at (-0.5, -1) {...};
		\node [style=Z phase dot] (10) at (-3, 0.875) {$\alpha$};
		\node [style=none] (11) at (-1, 1.625) {};
		\node [style=none] (12) at (-4, 1.625) {};
		\node [style=none] (13) at (-1, 0.375) {};
		\node [style=none, rotate=90] (14) at (-1, 1) {...};
		\node [style=none] (16) at (-4, 0.375) {};
		\node [style=none, rotate=90] (17) at (-3.75, 1) {...};
		\node [style=none] (18) at (-0.5, 0.375) {};
		\node [style=none] (19) at (-0.5, 1.625) {};
		\node [style=none] (22) at (-4, -1.625) {};
		\node [style=none] (23) at (-4, -0.375) {};
		\node [style=none] (24) at (0.5, 0.25) {$=$};
		\node [style=none, rotate=90] (25) at (-2.35, 0.05) {...};
		\node [style=none] (26) at (1.5, 1.25) {};
		\node [style=none, rotate=90] (28) at (2, 0) {...};
		\node [style=none] (29) at (5, -1.25) {};
		\node [style=none, rotate=90] (30) at (4.75, 0) {...};
		\node [style=Z phase dot] (32) at (3.25, 0) {$\ \alpha\!+\!\beta\ $};
		\node [style=none] (33) at (5, 1.25) {};
		\node [style=none] (34) at (1.5, -1.25) {};
	\end{pgfonlayer}
	\begin{pgfonlayer}{edgelayer}
		\draw [in=-141, out=0, looseness=0.75] (3.center) to (0);
		\draw [in=180, out=-39, looseness=0.75] (0) to (4.center);
		\draw [in=180, out=39, looseness=0.75] (0) to (1.center);
		\draw [in=180, out=0] (2.center) to (0);
		\draw [in=-141, out=0, looseness=0.75] (16.center) to (10);
		\draw [in=180, out=0] (10) to (13.center);
		\draw [in=180, out=39, looseness=0.75] (10) to (11.center);
		\draw [in=141, out=0, looseness=0.75] (12.center) to (10);
		\draw [bend left=45] (10) to (0);
		\draw (11.center) to (19.center);
		\draw (13.center) to (18.center);
		\draw (23.center) to (2.center);
		\draw (22.center) to (3.center);
		\draw [bend left=45] (0) to (10);
		\draw [in=-120, out=0] (34.center) to (32);
		\draw [in=180, out=-60] (32) to (29.center);
		\draw [in=180, out=60] (32) to (33.center);
		\draw [in=120, out=0] (26.center) to (32);
	\end{pgfonlayer}
\end{tikzpicture}}
\end{equation}


\section{IR Constructions of Benchmark Models}
\label{sec:IRBenMods}
In this section, we briefly introduce the models used for benchmarking, including their Hamiltonians and the corresponding IR construction.

    
    
    
    

\begin{figure}[h]
    \centering
    \includegraphics[width=0.34\textwidth]{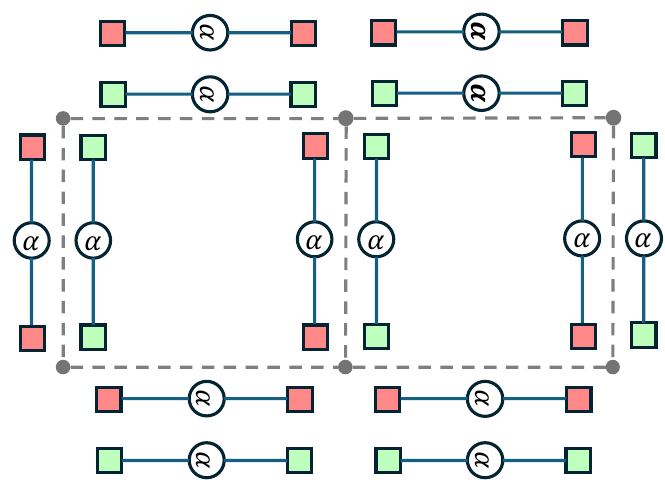}
    \caption{GadIR for spin-qubit model.}
    \label{fig:sqir}
\end{figure}

\subsection{Spin-Qubit Model}
Depicted in Figure~\ref{fig:sqir}, the \textit{Heisenberg model} is a well-known model for studying the phase transition of magnetic systems~\cite{heisenberg1928theorie,kittel2005int,troyer2005computational}, and it can be shown that the general Hubbard model introduced later can be reduced to it under appropriate assumptions~\cite{altland2006interaction,auerbach2012interacting,hubbard1963electron}. The Hamiltonian of the Heisenberg model can be written as
\begin{equation}
    H_{H}=\sum_{\langle i,j\rangle}(J_xX_iX_j+J_yY_iY_j+J_zZ_iZ_j)+\sum_ih_zZ_i,
\end{equation}
where $\langle i,j\rangle$ denotes the pair of neighboring sites; $J_x,\,J_y,\,J_z$ are the coupling strengths along each direction; $h_z$ is the magnetic field along the $z$ direction. 

If $J_z=0$, then the only interaction left is the coupling along the $x$ and $y$ directions, which gives rise to the Hamiltonian of the \textit{XY model}, $H_{XY}$.

\begin{figure}[h]
    \centering
    \includegraphics[width=0.32\textwidth]{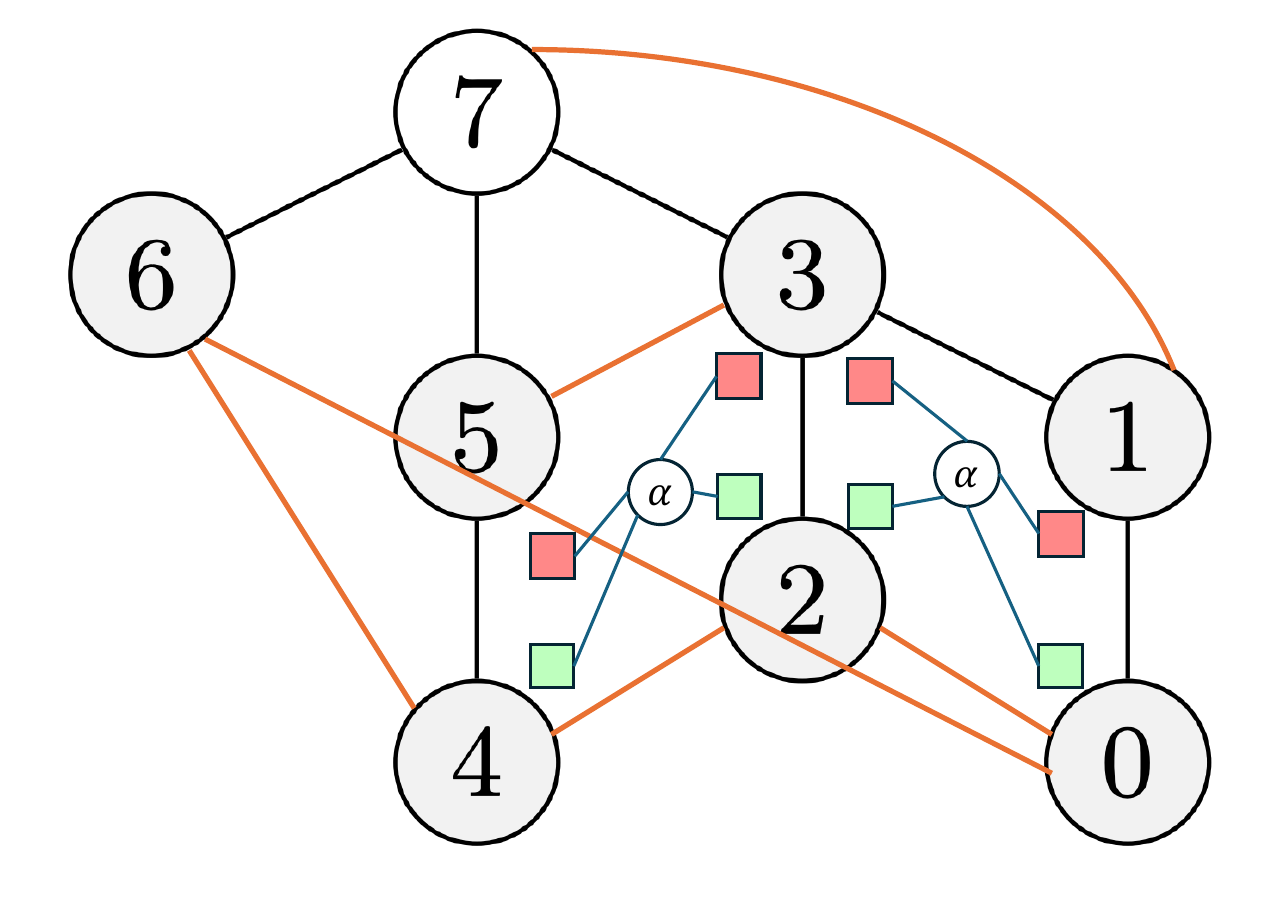}
    \caption{GadIR for Fermionic model.}
    \label{fig:fmir}
\end{figure}

\subsection{Fermion Model}
\subsubsection{Fermi-Hubbard}
 As one of the most famous and fundamental models, the \textit{Fermi-Hubbard model} is particularly useful in solid state physics to describe the crystal under the tight-binding approximation~\cite{altland2006interaction,hubbard1963electron}, which is closely related to the ultra-cold atoms in an optical lattice~\cite{bloch2008many,jordens2008mott}. A relatively general form of Hamiltonian for the Fermi-Hubbard model can be written as,
\begin{align}\notag
H_{FBS} = &-  t\sum_{\langle i, j\rangle } \sum_{\sigma}
             (c^\dagger_{i, \sigma} c_{j, \sigma} +
              c^\dagger_{j, \sigma} c_{i, \sigma})
    \\\notag
    &+ U
       \sum_{i} 
             n_{i,\uparrow} n_{i,\downarrow}
    \\
    &- \mu
       \sum_i \sum_{\sigma} n_{i, \sigma}
    - h \sum_{i} 
        \left(n_{i, \uparrow} - n_{i,\downarrow}\right),
\end{align}
where $c_{i,\sigma}$ is the fermionic field operator at site $i$, $n_{i,\sigma}=c_{i,\sigma}^\dagger c_{i,\sigma}$ is the number operator, and $n_i=\sum_{\sigma}n_{i,\sigma}$. $\langle i,j\rangle$ denotes a pair of neighboring sites;  $U$ is the Coulomb potential; $t$ is the tunneling amplitude; $\mu$ is the chemical potential; and $h$ is the magnetic field. 

To obtain the spinless version, one simply discards the spin degree of freedom $\sigma$ and the corresponding interaction, which arrives in 
\begin{align}\notag
H_{FB} = &-  t\sum_{\langle i, j\rangle }
             (c^\dagger_{i} c_{j} +
              c^\dagger_{j} c_{i})
    \\
    &+ U
       \sum_{\langle i,j\rangle} 
             n_{i} n_{j}
    - \mu\sum_i n_{i}.
\end{align}
Note that we will now have the Coulomb interaction between neighboring sites, which was previously dominated by the interaction on site and omitted in $H_{FBS}$.

\subsubsection{Jellium}
The \textit{Jellium model} is a special kind of the general Coulomb-governed interacting electron model under the Born-Oppenheimer approximation, where the background is idealized with a uniform positive charge density~\cite{giuliani2008quantum}. It is frequently applied for studying the fractional quantum Hall effect~\cite{laughlin1983anomalous}. The continuous Hamiltonian with $N$ electrons can be approximated as 
\begin{equation}
    \tilde{H}_{\mathrm{Jellium}}\approx\underbrace{\sum_i^N \frac{p_i^2}{2m}}_{T}+\underbrace{\sum_{i<j}^N\frac{e^2}{|\vec{r}_i-\vec{r}_j|}}_{V}.
\end{equation}
After choosing the plane-wave basis to discretize the model~\cite{babbush2018low}, the Hamiltonian has the following simple form,
\begin{equation}
    H_{\mathrm{Jellium}}=\sum_{pq}T_{pq}c^\dagger_pc_q+\sum_{pq}V_{pq}n_pn_q,
\end{equation}
with $p,q$ being the index of the site. $V_{pq}$ and $T_{pq}$ denote the kinetic energy strength and electron-electron repulsion, respectively, which can be calculated given the electron density. 

\begin{figure}[h]
    \centering
    \includegraphics[width=0.34\textwidth]{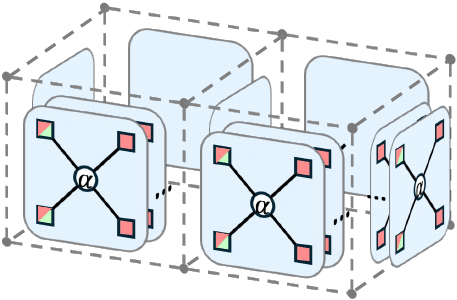}
    \caption{GadIR for Bosonic model.}
    \label{fig:bhir}
\end{figure}

\subsection{Boson Model}
Within the standard model, there are only two fundamental classes of particles concerning the particle-exchange statistic, namely bosons and fermions~\cite{weinberg1995quantum,griffiths2008introduction}. We also benchmark our proposed framework for simulating the dynamics of bosons, which may imply its universal feasibility and advantages across many-body models in physics, especially for nonlinear optics systems and superconducting systems~\cite{boyd2008nonlinear,blais2021circuit}.  The simulation of bosons is known to be more difficult than that of fermions, as the number of particles in a particular state is unlimited and unrestricted by the Pauli exclusion principle, which leads to infinite dimensions of the Hilbert space in the particle number basis and requires special truncation techniques such as Fock-state truncation~\cite{bloch2008many} or cluster expansion for Gaussian like states~\cite{kira2008cluster,huang2022classical,huang2023residual,plankensteiner2022quantumcumulants}.

In this work, we adopt the bosonic version of the Hubbard model~\cite{guo2012critical,ma1986strongly}, namely the Bose-Hubbard model, the Hamiltonian of which is written as 
\begin{align}\notag
H_{BH} = - t \sum_{\langle i, j \rangle} (b_i^\dagger b_j + b_j^\dagger b_i)
 + V \sum_{\langle i, j \rangle} b_i^\dagger b_i b_j^\dagger b_j \\
 + \frac{U}{2} \sum_i b_i^\dagger b_i (b_i^\dagger b_i - 1)
 - \mu \sum_i b_i^\dagger b_i.
\end{align}
where $b_i$ is the boson field operator; $t$ is now the hopping amplitude, $U$ is the on-site interaction; $\mu$ is the chemical potential. We use a relatively straightforward mapping method by mapping a site with maximum $n$ excitation to $n+1$ qubits, with each representing one of the excitation levels~\cite{miessen2021quantum}, in contrast to the one-to-one correspondence of site and qubit in the fermionic case. 

\begin{figure}[h]
    \centering
    \includegraphics[width=0.25\textwidth]{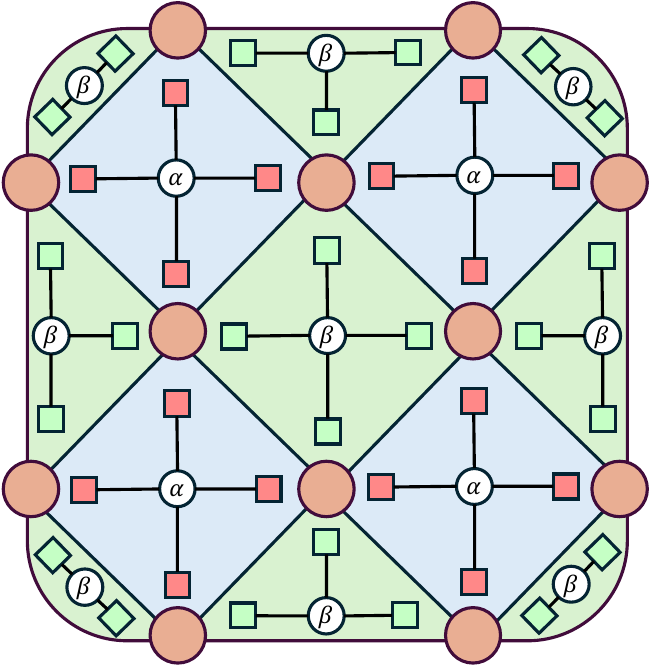}
    \caption{GadIR for lattice gauge theory model.}
    \label{fig:lgtir}
\end{figure}

\subsection{Lattice Gauge Theory}
Gauge theory constitutes the backbone of the standard model and thus also modern particle physics~\cite{weinberg1995quantum,griffiths2008introduction}. The Lagrangian, which can be transformed into a Hamiltonian governing the dynamics of the particles, can be derived directly from the symmetry group of the system, marking the milestone of theoretical physics. The gauge theory offers the prediction of the particle under all interactions except gravity that matches the experimental results to an ultra-high precision by modeling the interaction between matter fermions as a gauge field carried by the corresponding gauge bosons. Due to the complicated nature of the interaction, the dynamics is usually solved with perturbative methods organized with Feynman diagrams. However, this approach breaks down for quantum chromodynamics (QCD), which is itself a SU(3) gauge theory because of the famous color confinement~\cite{wilson1974confinement}, as the interaction becomes stronger as the energy decreases, and calls for a non-perturbative method.  Lattice gauge theory (LGT) is the most widely used method to overcome difficulties and provide fruitful results by discretizing the space, or in other words, working in a discretized position basis~\cite{gattringer2009quantum,rothe2012lattice}.
\begin{figure}[htb]
    \centering
    \includegraphics[width=0.4\textwidth]{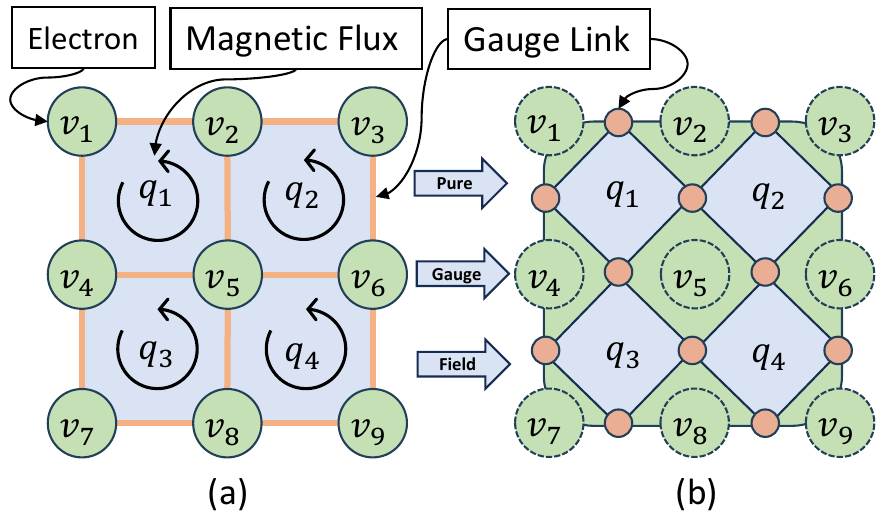}
    \caption{The illustration of the $\mathbb{Z}_2$ lattice gauge theory.}
    \label{fig:z2lgt}
\end{figure}

In this work, we strengthen the claim of our method's universal effectiveness by simulating a simplified version of quantum electrodynamics, where the $U(1)$ symmetry group is truncated into $\mathbb{Z}_2$~\cite{rothe2012lattice}. The matter fermions involved in this simplified model are electrons, while the gauge bosons are photons. The topology of the model can be represented in~\Cref{fig:z2lgt}(a), where the vertices represent the values of the electrons' wavefunction at a specific location of the discretized space, and the links represent the interaction (may also be considered the wavefunction) of the photons, which can be mapped to qubits. To further simplify the model, we can integrate out the wave function of the electrons and focus on the gauge field, which arrives at the topology represented in~\Cref{fig:z2lgt}(b) with each link mapped to a qubit. The Hamiltonian is, therefore, given by
\begin{align}\notag
    H_{LGT}=&-J_E\sum_{v_i}\prod_{l\in v_i}Z_l-J_M\sum_{p_i}\prod_{l\in p_i}X_p\\
            &-h_E\sum_lZ_l-\lambda\sum_lX_l.
\end{align}
Here, the first two terms serve as the 'detectors' for electrons and magnetic flux excitation with parameters $J_E$ and $J_M$, and the integral is performed over all the links involved in the vertices $v_i$ and plaquettes $p_i$ as shown in~\Cref{fig:z2lgt}(b), respectively. The last two terms denote electric fields and the hopping of the matter field introduced by the gauge field with the strengths $h_E$ and $\lambda$, respectively.

\subsection{Chemistry}
The Hamiltonian used for simulating the simple \textit{chemical molecules model} is a more general Coulomb Hamiltonian than the previously introduced Jellium model, which reads
\begin{equation}
H_{CY}=\sum_{pq} T_{pq}c_p^\dagger c_q+\sum_{pqrs} V_{pqrs} c_p^\dagger c_qc_r^\dagger c_s,
\end{equation}
under a general basis (instead of the plane wave one). Although the interaction part contains $N^4$ terms for simulating the molecules with $N$ sites (orbits), truncation of the eigenvalues with Cholesky decomposition can be applied to reduce the complexity to $O(N^2)$ as the rank of the matrix $V_{pqrs}$ generally scales linearly with $N$~\cite{motta2021low}. Given the desired number of active orbits (considering the spins, which double the number of qubits) and the geometry of the molecule, the parameters $T_{pq}$ and $V_{pqrs}$ can be calculated by the Colomb integral in the Molecular orbits with existing tools~\cite{mcclean2020openfermion}. And in this work, we select $\mathrm{LiH}$, $\mathrm{H_2O}$ and $\mathrm{Benzene}$ with $3,\,4,\,6$ active orbits, respectively, for benchmarks.

\section{Pseudo Code for Algorithm}\label{appendix:pseudo}
\renewcommand{\algorithmicwhile}{ \textbf{For}} 
\algdef{SE}[REPEATN]{RepeatN}{End}[1]{\algorithmicrepeat\ #1 \textbf{times}}{\algorithmicend}
\begin{algorithm}[h]
	\caption{Pauli Gadget Reduction-Tree Algorithm}
	\label{rand}
	\begin{algorithmic}[1]
		\Require Pauli gadget of model $P_m$, Topology $T$.
		\Ensure Gadget reduction tree $R$.
		\Function{Divide\_Group}{gadget sets $\{p\}$, Topology $T$}
                \While {edge $e$ in $T$}
                    \State Clifford set $G$, Reducible gadgets $\{p'\}$ $\gets$ iterate Clifford sets on $e$ and find maximal reducible \#gadgets
                \EndWhile
                \State $\{p_{rest}\} \gets $ $\{p\}/\{p'\}$
                \State \Return $\{p'\},\{p_{rest}\},G$
		\EndFunction
            
            \Function{Recursion\_Grow\_Children}{gadget sets $\{p\}$}
            \State Gadget set to divide $\{p_{div}\} \gets \{p\}$
                \While{True}
                    \State $\{p'\},\{p_{rest}\},G \gets$ \Call{Divide\_Group}{$\{p_{div}\},T$}
                    \If{$\{p'\}=$ None}
                        \State \Return
                    \EndIf
                    \State $\{p_{div}\} \gets \{p_{rest}\}$
                    \State Use $G$ to apply group reduction on $\{p'\}$
                    \State $\{p_{reduced}\} \gets$ reduce gadgets from $\{p'\}$
                    \State Node of $\{p\}$ grow its child as node of $\{p_{reduced}\}$
                \EndWhile
		\EndFunction
            
            \State $R \gets $ \Call{Recursion\_Grow\_Children}{$P_m$}
	\end{algorithmic}
    \label{alg:reductiontree}
\end{algorithm}

\clearpage
\section{Pauli Gadget Reduction Rules}
\label{appendix:reductionRules}

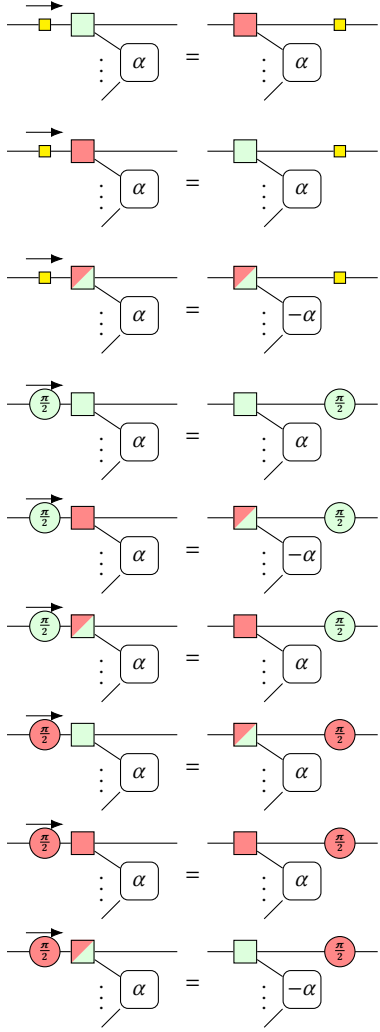
\begin{figure}[h!]
    \centering
    \begin{equation}\nonumber
    {\begin{tikzpicture}[baseline={(16)}]
    \draw[-{Latex}]        (0.5,0.5)   -- (1.5,0.5);
	\begin{pgfonlayer}{nodelayer}
		\node [style=none] (1) at (0, 0) {};
		\node [style=none] (7) at (2.5, -2) {};
		\node [style=none] (9) at (4.5, 0) {};
		\node [style=none] (10) at (2.5, -1) {$\vdots$};
		\node [style=hadamard] (14) at (1, 0) {};
		\node [style=Zexp] (15) at (2, 0) {};
		\node [style=Aexp] (16) at (3.5, -1) {$\alpha$};
	\end{pgfonlayer}
	\begin{pgfonlayer}{edgelayer}
		\draw (1.center) to (14);
		\draw (14) to (15);
		\draw (15) to (9.center);
		\draw (16) to (7.center);
		\draw (16) to (15);
	\end{pgfonlayer}
\end{tikzpicture}} = {\begin{tikzpicture}[baseline={(16)}]
	\begin{pgfonlayer}{nodelayer}
		\node [style=none] (1) at (0, 0) {};
		\node [style=none] (7) at (1.5, -2) {};
		\node [style=none] (9) at (4.5, 0) {};
		\node [style=none] (10) at (1.5, -1) {$\vdots$};
		\node [style=hadamard] (14) at (3.5, 0) {};
		\node [style=Xexp] (15) at (1, 0) {};
		\node [style=Aexp] (16) at (2.5, -1) {$\alpha$};
	\end{pgfonlayer}
	\begin{pgfonlayer}{edgelayer}
		\draw (1.center) to (15);
		\draw (14) to (9.center);
		\draw (14) to (15);
		\draw (16) to (15);
		\draw (16) to (7.center);
	\end{pgfonlayer}
\end{tikzpicture}}
    \end{equation}
    
    \begin{equation}\nonumber
    {\begin{tikzpicture}[baseline={(16)}]
\draw[-{Latex}]        (0.5,0.5)   -- (1.5,0.5);
	\begin{pgfonlayer}{nodelayer}
		\node [style=none] (1) at (0, 0*2) {};
		\node [style=none] (7) at (1.25*2, -1*2) {};
		\node [style=none] (9) at (2.25*2, 0*2) {};
		\node [style=none] (10) at (1.25*2, -0.5*2) {$\vdots$};
		\node [style=hadamard] (14) at (0.5*2, 0*2) {};
		\node [style=Xexp] (15) at (1*2, 0*2) {};
		\node [style=Aexp] (16) at (1.75*2, -0.5*2) {$\alpha$};
	\end{pgfonlayer}
	\begin{pgfonlayer}{edgelayer}
		\draw (1.center) to (14);
		\draw (14) to (15);
		\draw (15) to (9.center);
		\draw (16) to (7.center);
		\draw (16) to (15);
	\end{pgfonlayer}
\end{tikzpicture}} = {\begin{tikzpicture}[baseline={(16)}]
	\begin{pgfonlayer}{nodelayer}
		\node [style=none] (1) at (0, 0) {};
		\node [style=none] (7) at (0.75*2, -1*2) {};
		\node [style=none] (9) at (2.25*2, 0) {};
		\node [style=none] (10) at (0.75*2, -0.5*2) {$\vdots$};
		\node [style=hadamard] (14) at (1.75*2, 0) {};
		\node [style=Zexp] (15) at (0.5*2, 0) {};
		\node [style=Aexp] (16) at (1.25*2, -0.5*2) {$\alpha$};
	\end{pgfonlayer}
	\begin{pgfonlayer}{edgelayer}
		\draw (1.center) to (15);
		\draw (14) to (9.center);
		\draw (14) to (15);
		\draw (16) to (15);
		\draw (16) to (7.center);
	\end{pgfonlayer}
\end{tikzpicture}}
    \end{equation}
    
    \begin{equation}\nonumber
    {\begin{tikzpicture}[baseline={(16)}]
\draw[-{Latex}]        (0.5,0.5)   -- (1.5,0.5);
	\begin{pgfonlayer}{nodelayer}
		\node [style=none] (1) at (0, 0) {};
		\node [style=none] (7) at (1.25*2, -1*2) {};
		\node [style=none] (9) at (2.25*2, 0) {};
		\node [style=none] (10) at (1.25*2, -0.5*2) {$\vdots$};
		\node [style=hadamard] (14) at (0.5*2, 0) {};
		\node [style=Yexp] (15) at (1*2, 0) {};
		\node [style=Aexp] (16) at (1.75*2, -0.5*2) {$\alpha$};
	\end{pgfonlayer}
	\begin{pgfonlayer}{edgelayer}
		\draw (1.center) to (14);
		\draw (14) to (15);
		\draw (15) to (9.center);
		\draw (16) to (7.center);
		\draw (16) to (15);
	\end{pgfonlayer}
\end{tikzpicture}} = {\begin{tikzpicture}[baseline={(16)}]
	\begin{pgfonlayer}{nodelayer}
		\node [style=none] (1) at (0, 0) {};
		\node [style=none] (7) at (0.75*2, -1*2) {};
		\node [style=none] (9) at (2.25*2, 0) {};
		\node [style=none] (10) at (0.75*2, -0.5*2) {$\vdots$};
		\node [style=hadamard] (14) at (1.75*2, 0) {};
		\node [style=Yexp] (15) at (0.5*2, 0) {};
		\node [style=Aexp] (16) at (1.25*2, -0.5*2) {$-\alpha$};
	\end{pgfonlayer}
	\begin{pgfonlayer}{edgelayer}
		\draw (1.center) to (15);
		\draw (14) to (9.center);
		\draw (14) to (15);
		\draw (16) to (15);
		\draw (16) to (7.center);
	\end{pgfonlayer}
\end{tikzpicture}}
    \end{equation}
    
    \begin{equation}\nonumber
    {\begin{tikzpicture}[baseline={(16)}]
\draw[-{Latex}]        (0.5,0.5)   -- (1.5,0.5);
	\begin{pgfonlayer}{nodelayer}
		\node [style=none] (1) at (0, 0) {};
		\node [style=none] (7) at (1.25*2, -1*2) {};
		\node [style=none] (9) at (2.25*2, 0) {};
		\node [style=none] (10) at (1.25*2, -0.5*2) {$\vdots$};
		\node [style=Z phase dot] (14) at (0.5*2, 0) {$\frac{\pi}{2}$};
		\node [style=Zexp] (15) at (1*2, 0) {};
		\node [style=Aexp] (16) at (1.75*2, -0.5*2) {$\alpha$};
	\end{pgfonlayer}
	\begin{pgfonlayer}{edgelayer}
		\draw (1.center) to (14);
		\draw (14) to (15);
		\draw (15) to (9.center);
		\draw (16) to (7.center);
		\draw (16) to (15);
	\end{pgfonlayer}
\end{tikzpicture}} = {\begin{tikzpicture}[baseline={(16)}]
	\begin{pgfonlayer}{nodelayer}
		\node [style=none] (1) at (0, 0) {};
		\node [style=none] (7) at (0.75*2, -1*2) {};
		\node [style=none] (9) at (2.25*2, 0) {};
		\node [style=none] (10) at (0.75*2, -0.5*2) {$\vdots$};
		\node [style=Z phase dot] (14) at (1.75*2, 0) {$\frac{\pi}{2}$};
		\node [style=Zexp] (15) at (0.5*2, 0) {};
		\node [style=Aexp] (16) at (1.25*2, -0.5*2) {$\alpha$};
	\end{pgfonlayer}
	\begin{pgfonlayer}{edgelayer}
		\draw (1.center) to (15);
		\draw (14) to (9.center);
		\draw (14) to (15);
		\draw (16) to (15);
		\draw (16) to (7.center);
	\end{pgfonlayer}
\end{tikzpicture}}
    \end{equation}
    \begin{equation}\nonumber
    {\begin{tikzpicture}[baseline={(16)}]
    \draw[-{Latex}]        (0.5,0.5)   -- (1.5,0.5);
	\begin{pgfonlayer}{nodelayer}
		\node [style=none] (1) at (0, 0) {};
		\node [style=none] (7) at (2.5, -2) {};
		\node [style=none] (9) at (4.5, 0) {};
		\node [style=none] (10) at (2.5, -1) {$\vdots$};
		\node [style=Z phase dot] (14) at (1, 0) {$\frac{\pi}{2}$};
		\node [style=Xexp] (15) at (2, 0) {};
		\node [style=Aexp] (16) at (3.5, -1) {$\alpha$};
	\end{pgfonlayer}
	\begin{pgfonlayer}{edgelayer}
		\draw (1.center) to (14);
		\draw (14) to (15);
		\draw (15) to (9.center);
		\draw (16) to (7.center);
		\draw (16) to (15);
	\end{pgfonlayer}
\end{tikzpicture}} = {\begin{tikzpicture}[baseline={(16)}]
	\begin{pgfonlayer}{nodelayer}
		\node [style=none] (1) at (0, 0) {};
		\node [style=none] (7) at (1.5, -2) {};
		\node [style=none] (9) at (4.5, 0) {};
		\node [style=none] (10) at (1.5, -1) {$\vdots$};
		\node [style=Z phase dot] (14) at (3.5, 0) {$\frac{\pi}{2}$};
		\node [style=Yexp] (15) at (1, 0) {};
		\node [style=Aexp] (16) at (2.5, -1) {$-\alpha$};
	\end{pgfonlayer}
	\begin{pgfonlayer}{edgelayer}
		\draw (1.center) to (15);
		\draw (14) to (9.center);
		\draw (14) to (15);
		\draw (16) to (15);
		\draw (16) to (7.center);
	\end{pgfonlayer}
\end{tikzpicture}}
    \end{equation}
    \begin{equation}\nonumber
    {\begin{tikzpicture}[baseline={(16)}]
\draw[-{Latex}]        (0.5,0.5)   -- (1.5,0.5);
	\begin{pgfonlayer}{nodelayer}
		\node [style=none] (1) at (0, 0) {};
		\node [style=none] (7) at (1.25*2, -1*2) {};
		\node [style=none] (9) at (2.25*2, 0) {};
		\node [style=none] (10) at (1.25*2, -0.5*2) {$\vdots$};
		\node [style=Z phase dot] (14) at (0.5*2, 0) {$\frac{\pi}{2}$};
		\node [style=Yexp] (15) at (1*2, 0) {};
		\node [style=Aexp] (16) at (1.75*2, -0.5*2) {$\alpha$};
	\end{pgfonlayer}
	\begin{pgfonlayer}{edgelayer}
		\draw (1.center) to (14);
		\draw (14) to (15);
		\draw (15) to (9.center);
		\draw (16) to (7.center);
		\draw (16) to (15);
	\end{pgfonlayer}
\end{tikzpicture}} = {\begin{tikzpicture}[baseline={(16)}]
	\begin{pgfonlayer}{nodelayer}
		\node [style=none] (1) at (0, 0) {};
		\node [style=none] (7) at (0.75*2, -1*2) {};
		\node [style=none] (9) at (2.25*2, 0) {};
		\node [style=none] (10) at (0.75*2, -0.5*2) {$\vdots$};
		\node [style=Z phase dot] (14) at (1.75*2, 0) {$\frac{\pi}{2}$};
		\node [style=Xexp] (15) at (0.5*2, 0) {};
		\node [style=Aexp] (16) at (1.25*2, -0.5*2) {$\alpha$};
	\end{pgfonlayer}
	\begin{pgfonlayer}{edgelayer}
		\draw (1.center) to (15);
		\draw (14) to (9.center);
		\draw (14) to (15);
		\draw (16) to (15);
		\draw (16) to (7.center);
	\end{pgfonlayer}
\end{tikzpicture}}
    \end{equation}
    \begin{equation}\nonumber
    {\begin{tikzpicture}[baseline={(16)}]
\draw[-{Latex}]        (0.5,0.5)   -- (1.5,0.5);
	\begin{pgfonlayer}{nodelayer}
		\node [style=none] (1) at (0, 0) {};
		\node [style=none] (7) at (1.25*2, -1*2) {};
		\node [style=none] (9) at (2.25*2, 0) {};
		\node [style=none] (10) at (1.25*2, -0.5*2) {$\vdots$};
		\node [style=X phase dot] (14) at (0.5*2, 0) {$\frac{\pi}{2}$};
		\node [style=Zexp] (15) at (1*2, 0) {};
		\node [style=Aexp] (16) at (1.75*2, -0.5*2) {$\alpha$};
	\end{pgfonlayer}
	\begin{pgfonlayer}{edgelayer}
		\draw (1.center) to (14);
		\draw (14) to (15);
		\draw (15) to (9.center);
		\draw (16) to (7.center);
		\draw (16) to (15);
	\end{pgfonlayer}
\end{tikzpicture}} = {\begin{tikzpicture}[baseline={(16)}]
	\begin{pgfonlayer}{nodelayer}
		\node [style=none] (1) at (0, 0) {};
		\node [style=none] (7) at (0.75*2, -1*2) {};
		\node [style=none] (9) at (2.25*2, 0) {};
		\node [style=none] (10) at (0.75*2, -0.5*2) {$\vdots$};
		\node [style=X phase dot] (14) at (1.75*2, 0) {$\frac{\pi}{2}$};
		\node [style=Yexp] (15) at (0.5*2, 0) {};
		\node [style=Aexp] (16) at (1.25*2, -0.5*2) {$\alpha$};
	\end{pgfonlayer}
	\begin{pgfonlayer}{edgelayer}
		\draw (1.center) to (15);
		\draw (14) to (9.center);
		\draw (14) to (15);
		\draw (16) to (15);
		\draw (16) to (7.center);
	\end{pgfonlayer}
\end{tikzpicture}}
    \end{equation}
    \begin{equation}\nonumber
    {\begin{tikzpicture}[baseline={(16)}]
\draw[-{Latex}]        (0.5,0.5)   -- (1.5,0.5);
	\begin{pgfonlayer}{nodelayer}
		\node [style=none] (1) at (0, 0) {};
		\node [style=none] (7) at (1.25*2, -1*2) {};
		\node [style=none] (9) at (2.25*2, 0) {};
		\node [style=none] (10) at (1.25*2, -0.5*2) {$\vdots$};
		\node [style=X phase dot] (14) at (0.5*2, 0) {$\frac{\pi}{2}$};
		\node [style=Xexp] (15) at (1*2, 0) {};
		\node [style=Aexp] (16) at (1.75*2, -0.5*2) {$\alpha$};
	\end{pgfonlayer}
	\begin{pgfonlayer}{edgelayer}
		\draw (1.center) to (14);
		\draw (14) to (15);
		\draw (15) to (9.center);
		\draw (16) to (7.center);
		\draw (16) to (15);
	\end{pgfonlayer}
\end{tikzpicture}} = {\begin{tikzpicture}[baseline={(16)}]
	\begin{pgfonlayer}{nodelayer}
		\node [style=none] (1) at (0, 0) {};
		\node [style=none] (7) at (0.75*2, -1*2) {};
		\node [style=none] (9) at (2.25*2, 0) {};
		\node [style=none] (10) at (0.75*2, -0.5*2) {$\vdots$};
		\node [style=X phase dot] (14) at (1.75*2, 0) {$\frac{\pi}{2}$};
		\node [style=Xexp] (15) at (0.5*2, 0) {};
		\node [style=Aexp] (16) at (1.25*2, -0.5*2) {$\alpha$};
	\end{pgfonlayer}
	\begin{pgfonlayer}{edgelayer}
		\draw (1.center) to (15);
		\draw (14) to (9.center);
		\draw (14) to (15);
		\draw (16) to (15);
		\draw (16) to (7.center);
	\end{pgfonlayer}
\end{tikzpicture}}
    \end{equation}
    \begin{equation}\nonumber
    {\begin{tikzpicture}[baseline={(16)}]
\draw[-{Latex}]        (0.5,0.5)   -- (1.5,0.5);
	\begin{pgfonlayer}{nodelayer}
		\node [style=none] (1) at (0, 0) {};
		\node [style=none] (7) at (1.25*2, -1*2) {};
		\node [style=none] (9) at (2.25*2, 0) {};
		\node [style=none] (10) at (1.25*2, -0.5*2) {$\vdots$};
		\node [style=X phase dot] (14) at (0.5*2, 0) {$\frac{\pi}{2}$};
		\node [style=Yexp] (15) at (1*2, 0) {};
		\node [style=Aexp] (16) at (1.75*2, -0.5*2) {$\alpha$};
	\end{pgfonlayer}
	\begin{pgfonlayer}{edgelayer}
		\draw (1.center) to (14);
		\draw (14) to (15);
		\draw (15) to (9.center);
		\draw (16) to (7.center);
		\draw (16) to (15);
	\end{pgfonlayer}
\end{tikzpicture}} = {\begin{tikzpicture}[baseline={(16)}]
	\begin{pgfonlayer}{nodelayer}
		\node [style=none] (1) at (0, 0) {};
		\node [style=none] (7) at (0.75*2, -1*2) {};
		\node [style=none] (9) at (2.25*2, 0) {};
		\node [style=none] (10) at (0.75*2, -0.5*2) {$\vdots$};
		\node [style=X phase dot] (14) at (1.75*2, 0) {$\frac{\pi}{2}$};
		\node [style=Zexp] (15) at (0.5*2, 0) {};
		\node [style=Aexp] (16) at (1.25*2, -0.5*2) {$-\alpha$};
	\end{pgfonlayer}
	\begin{pgfonlayer}{edgelayer}
		\draw (1.center) to (15);
		\draw (14) to (9.center);
		\draw (14) to (15);
		\draw (16) to (15);
		\draw (16) to (7.center);
	\end{pgfonlayer}
\end{tikzpicture}}
    \end{equation}
    \caption{Clifford gate passing through rules for gadgets.}
    \label{fig:cfpr}
\end{figure}

\begin{figure*}[htb]
    \centering
    \includegraphics[width=\textwidth]{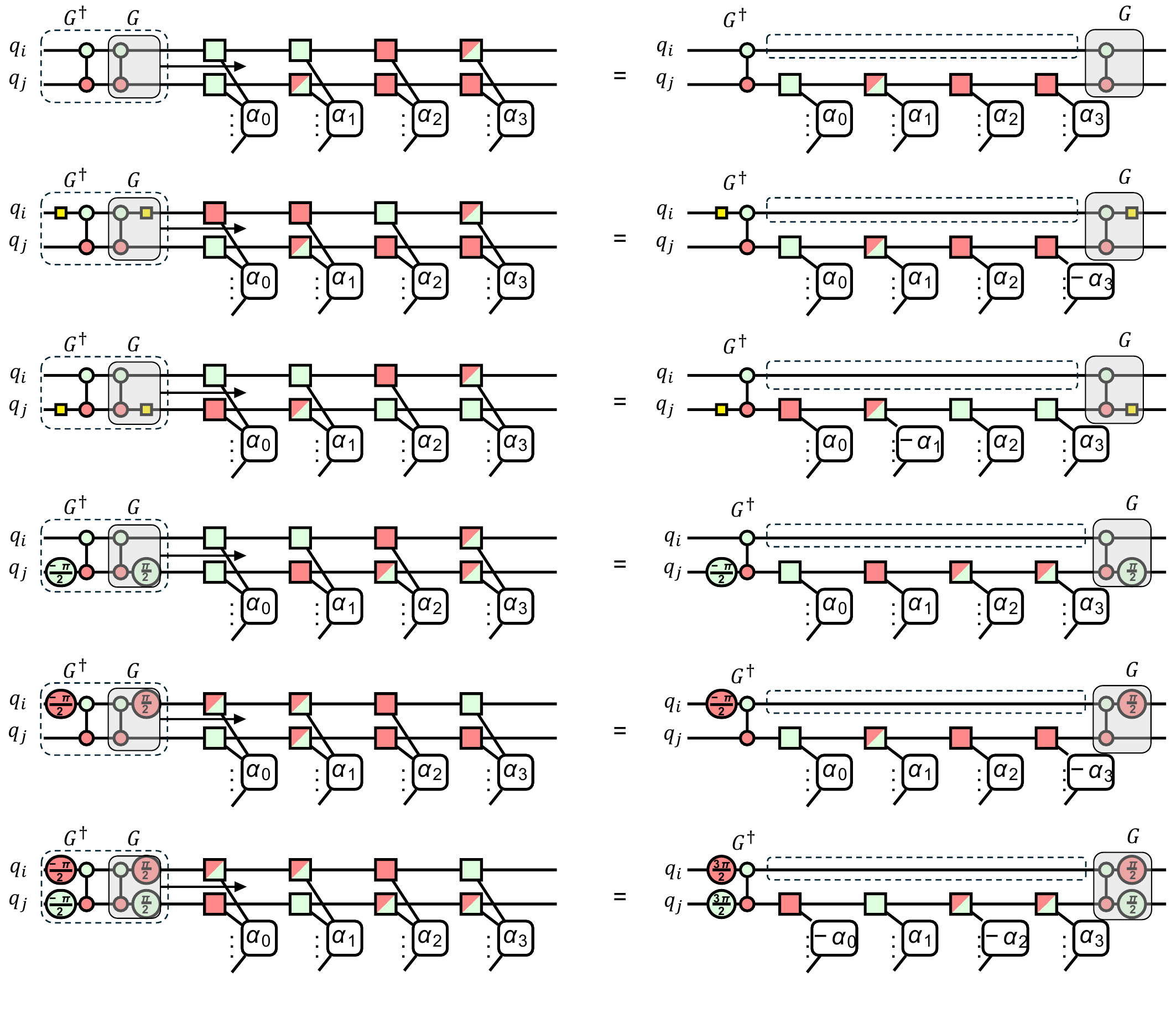}
    \caption{Group reduction rules for Pauli gadgets.}
    \label{fig:grdr}
\end{figure*}

\clearpage
%
%

\appendix
\section{Artifact Appendix}

\subsection{Abstract}

This artifact reproduces the GadIR evaluation in Figures~10--15, 17, and
18(d).  It runs GadIR and fixed baselines on superconducting
(Figures~10--11), neutral-atom (Figure~12), FTQC (Figures~13--14), and MBQC
(Figure~15) backends.  It also evaluates group reduction on random
Hamiltonians (Figure~17) and MPS entanglement scaling (Figure~18(d)).
The archive includes the generated result tables and figures for immediate
plot-only validation.  No quantum hardware is required.

\subsection{Artifact check-list (meta-information)}
{\small
\begin{itemize}
  \item {\bf Algorithm: } GadIR and fixed baselines use one second-order
  Suzuki step in Figures~10--15; Figure~17 is a topology-agnostic
  group-reduction ablation.
  \item {\bf Program: } Python architecture drivers and packaged QuCLEAR,
  2QAN, Tetris, Rustiq, PyZX, ZAC, and GSsynth code.  TQEC is installed
  from a pinned upstream Git revision.
  \item {\bf Compilation: } Qiskit synthesis/routing; a Rust extension for
  Rustiq.
  \item {\bf Transformations: } Pauli-gadget reduction, routing/scheduling,
  FTQC conversion/simulation, and MBQC graph synthesis.
  \item {\bf Binary: } Rustiq is built locally; the remaining Python code
  runs from packaged source or pinned installed dependencies.
  \item {\bf Model: } 52 Hamiltonians from eight physical/model families,
  plus random Hamiltonians for the group-reduction ablation.
  \item {\bf Data set: } JSON Hamiltonians and hardware specifications.
  \item {\bf Run-time environment: } Three pinned Conda environments.  The
  analysis environment uses Python~3.10 and Qiskit~1.4.3; QuCLEAR uses
  Python~3.10 and Qiskit~1.1.1; 2QAN uses Python~3.9 and Qiskit~0.36.2.
  \item {\bf Hardware: } CPU-only; at least 16 Intel, AMD, or Apple CPU cores,
  32\,GB RAM, and 20\,GB disk recommended.
  \item {\bf Run-time state: } Generated tables and figures are included
  under \path{results/} and \path{output/}; full reproduction overwrites them
  with fresh outputs.
  \item {\bf Execution: } One shell entry point; four architecture backends
  followed by the analysis workflows.
  \item {\bf Metrics: } Two-qubit count, depth, duration, fidelity,
  non-Clifford count, logical error, emitter-entanglement count, reduction
  percentage, and maximum MPS bipartite entropy.
  \item {\bf Output: } CSV/QASM/statistics and ten PDF files for eight
  numbered figures, including full-range variants of Figures~10 and~11.
  \item {\bf Experiments: } Smoke, full reproduction, and plot-only modes.
  \item {\bf How much disk space required (approximately)?: } 20\,GB.
  \item {\bf How much time is needed to prepare workflow (approximately)?: }
  30--90 minutes, depending on network/build speed.
  \item {\bf How much time is needed to complete experiments
  (approximately)?: } Smoke: tens of minutes; full run: hours to multiple
  days.
  \item {\bf Publicly available?: } Yes, through GitHub and Zenodo.
  \item {\bf Code licenses (if publicly available)?: } This AE folder does
  not introduce a GadIR license.  Redistributed third-party code retains its
  upstream notice in \path{baselines/<name>/LICENSE}; TQEC is pinned in the
  analysis environment specification.
  \item {\bf Data licenses (if publicly available)?: } See the released
  artifact and Zenodo record.
  \item {\bf Workflow automation framework used?: } Bash and Python.
  \item {\bf Archived (provide DOI)?: }
  \url{https://doi.org/10.5281/zenodo.21550065}.
\end{itemize}
}

\subsection{Description}

\subsubsection{How to access}

Repository: \url{https://github.com/xiyurain/GadIR}.

Archive: \url{https://doi.org/10.5281/zenodo.21550065}.

The public repository can be cloned over HTTPS, and the archive can be used
without GitHub access.  All commands run from \path{GadIR_AE/}.

\subsubsection{Hardware dependencies}

No quantum processor or GPU is required.  Intel, AMD, and Apple CPUs are
supported; we recommend at least 16 CPU cores, 32\,GB RAM, and 20\,GB disk.

\subsubsection{Software dependencies}

The host needs Bash, Git, Conda, C/C++ build tools, stable Rust/Cargo, and
network access during installation.  \path{environments/} pins:
\begin{itemize}
  \item \path{analysis.yml}: Python~3.10, Qiskit~1.4.3,
  Qiskit Aer~0.17.1, PyZX~0.9.0, scientific/plotting packages,
  Stim/sinter~1.15.0, PyMatching~2.3.0, pinned TQEC, and Jupyter Notebook;
  \item \path{quclear.yml}: Python~3.10, Qiskit~1.1.1, and pytest~7.4.4.
  \item \path{2qan.yml}: Python~3.9 and the pinned build/runtime
  dependencies for Qiskit Terra~0.20.2 (the Terra release in Qiskit~0.36.2).
\end{itemize}
2QAN, QuCLEAR, Tetris, Rustiq, PyZX, and ZAC code used by the scripts is under
\path{baselines/}.  The Python GSsynth Naive implementation is under
\path{src/mbqc/}, with its upstream GPL-3.0 license under
\path{baselines/gssynth/}.  Conda installs the exact TQEC revision pinned in
\path{analysis.yml}.  The bundled GraphiQ compatibility wheel is installed
without its obsolete Qiskit dependency, but the Figure~15 workflow uses
GSsynth Naive.  Packaged upstream licenses and available revision records are
stored with the corresponding baseline.

\subsubsection{Data sets}

There are 52 JSON Hamiltonians under \path{benchmarks/hamiltonians/} and
architectures under \path{hardware/}.

\subsubsection{Models}

The compiler evaluations in Figures~10--15 use one second-order Suzuki step.
Figure~13 uses eight models, Figure~14 uses FH(4,5)/BH(4,5), and Figure~15
uses 21 models, excluding only \path{jellium_3_3}.  Figure~17 compares
prior individual-gadget reduction with GadIR group reduction using 1000 random
instances at each Hamiltonian-term count and fixed seed~7; this ablation is
topology-agnostic.
Figure~18(d) covers Heisenberg, Bose--Hubbard, Fermi--Hubbard, Jellium, and
molecule model families.  It combines exact profiles computed from packaged
Hamiltonians with dimension-derived bounds.  Exact profiles use four
product-formula steps at evolution times 0.4, 0.8, and 1.0.  Because the
structural Hamiltonian inputs do not contain numerical coefficients, every
term uses coefficient~1.

\subsection{Installation}

The following numbered procedure mirrors the explicit environment setup used
by the QuCLEAR artifact and should be followed on a fresh machine.

\begin{enumerate}
  \item Install Miniconda or Anaconda from
  \url{https://docs.conda.io/projects/miniconda/}, initialize it for the
  current shell, open a new shell, and check:
\begin{verbatim}
conda --version
git --version
\end{verbatim}

  \item Install the host C/C++ build tools using the operating system package
  manager.  Install stable Rust and Cargo using the official instructions at
  \url{https://www.rust-lang.org/tools/install}; the standard rustup command is:
\begin{verbatim}
curl --proto '=https' --tlsv1.2 -sSf \
  https://sh.rustup.rs | sh
source "$HOME/.cargo/env"
rustc --version
cargo --version
\end{verbatim}

  \item Clone the repository and enter the self-contained artifact directory:
\begin{verbatim}
git clone https://github.com/xiyurain/GadIR.git
cd GadIR/GadIR_AE
\end{verbatim}

  \item Create all environments and build/install the packaged Rustiq
  component with the single installation script:
\begin{verbatim}
./prepare.sh
\end{verbatim}
This creates \path{.envs/analysis}, \path{.envs/quclear}, and
\path{.envs/2qan} from the three YAML files described above.  It also builds
and installs the Rustiq Python extension from \path{baselines/rustiq/} into
the analysis environment.  The run scripts select the appropriate interpreter
automatically, so manual environment activation is unnecessary.
Tweedledum~1.1.1 is installed without build isolation before Qiskit
Terra~0.20.2 in the 2QAN environment, using the pinned CMake/scikit-build
toolchain from \path{2qan.yml}.

  \item Verify the three environments before starting a long run:
\begin{verbatim}
./.envs/analysis/bin/python -c \
 "import qiskit,pyzx,rustiq,stim,sinter"
./.envs/analysis/bin/python -c \
 "import pymatching,tqec"
./.envs/quclear/bin/python -c \
 "import qiskit"
./.envs/2qan/bin/python -c \
 "import qiskit"
\end{verbatim}
These commands verify the installed dependencies.  The next section first
checks the bundled results and then runs the functional smoke test.
\end{enumerate}

\subsection{Experiment workflow}

The included tables should first be checked without rerunning the experiments:
\begin{verbatim}
./run.sh plot
\end{verbatim}
\path{plot} redraws every PDF from the included tables.  Next, run the reduced
end-to-end validation across all four architecture backends and the analysis
workflows:
\begin{verbatim}
./run.sh smoke
\end{verbatim}
The smoke test writes reduced result tables and figures, so the plot-only check
must precede it.  The complete, potentially long reproduction restores the
full tables and figures from the packaged Hamiltonians:
\begin{verbatim}
./run.sh reproduce
\end{verbatim}
This command overwrites the bundled tables under \path{results/} and
regenerates Figures~10--15, 17, and 18(d).  The two analysis workflows write
\path{results/analysis/figure17.csv},
\path{results/analysis/figure18d.csv}, \path{output/figure17.pdf}, and
\path{output/figure18d.pdf}.

To generate all paper figures from completed AE results, use the provided
plot-only notebook:
\begin{verbatim}
./.envs/analysis/bin/jupyter notebook \
  generate_all_figures.ipynb
\end{verbatim}
It draws Figures~10--15, 17, and 18(d) into \path{output/}; run
\path{./run.sh reproduce} first when freshly regenerated rather than bundled
tables are required.

The expected PDFs are:
\begin{verbatim}
output/figure10.pdf
output/figure10_full_range.pdf
output/figure11.pdf
output/figure11_full_range.pdf
output/figure12.pdf
output/figure13.pdf
output/figure14.pdf
output/figure15.pdf
output/figure17.pdf
output/figure18d.pdf
\end{verbatim}

Workflows run in the fixed order superconducting, neutral atom, FTQC, MBQC,
and analysis.

\subsection{Evaluation and expected results}

A successful run places raw data in \path{results/<backend>/} and PDFs in
\path{output/}.  Figure~10 reports normalized IBM CNOT/Google CZ count;
Figure~11, normalized depth; Figure~12, neutral-atom two-qubit count, duration,
and fidelity; Figure~13, QuCLEAR/GadIR non-Clifford count; Figure~14, fresh
Stim/sinter statistics and logical-CNOT error extrapolation; and Figure~15,
Tetris/QuCLEAR/PauliEvo-Rustiq/GadIR emitter-entanglement count.  Figure~15
uses the same Suzuki-2 construction for all four frontends.  QuCLEAR and
PauliEvo-Rustiq optimize the forward and reverse halves independently; no
Qiskit O3 pass is applied.  Every method is reduced by PyZX
\path{clifford_simp} and synthesized by the packaged Python GSsynth Naive
backend.  GadIR includes the cost of reconnecting its independently compiled
leaves.  Figure~17 reports, for each
term count, the mean reduction percentage over 1000 random instances for
individual- and group-reduction methods.  Figure~18(d) reports maximum
bipartite entropy over MPS prefix cuts.  Solid markers are exact profiles from
packaged Hamiltonians, and open markers are dimension-derived bounds.
Figure~14 uses stochastic Stim/sinter sampling and may show small run-to-run
variation; the other workflows use fixed seeds or deterministic settings.
The regenerated tables and figures should match the revised paper figures.
For Figures~10--11, 2QAN runs one fixed seed on the Heisenberg and XY models
that have 2QAN points in the paper.  Each physical edge is represented by one
structural gate whose body retains separate XX/YY/ZZ or XX/YY rotations.
GadIR configurations are fixed by architecture and model family, and are
shared by all sizes within a family.  Baseline configurations are fixed and
do not use per-benchmark trial selection.  Figure~12 uses one calibrated noise
model for all methods, and its GadIR candidate is also fixed by model family.
Figure~14 is an extrapolation rather than full non-Clifford physical
simulation; Figure~15 uses one deterministic trial and one uniform backend,
without a fallback.

\subsection{Experiment customization}

For diagnosis, a backend may be run as:
\begin{verbatim}
./.envs/analysis/bin/python -m src.<backend>.run smoke
\end{verbatim}
Changing fixed configurations is outside the reported evaluation.

\subsection{Methodology}

Submission, reviewing, and badging methodology:
\begin{itemize}
  \item \url{https://www.acm.org/publications/policies/artifact-review-and-badging-current}
  \item \url{https://cTuning.org/ae}
\end{itemize}

\end{document}